\documentclass[fleqn,usenatbib]{mnras}
\usepackage{newtxtext,newtxmath}

\usepackage[T1]{fontenc}

\DeclareRobustCommand{\VAN}[3]{#2}
\let\VANthebibliography\thebibliography
\def\thebibliography{\DeclareRobustCommand{\VAN}[3]{##3}\VANthebibliography}

\usepackage{graphicx}	% Including figure files
\usepackage{amsmath}	% Advanced maths commands
\usepackage{orcidlink}
\usepackage{hyperref}
\usepackage{braket}
\usepackage{cuted}
\usepackage{widetext}
\title[Self-calibration of IA in DES year 1]{Extracting intrinsic alignments in the Dark Energy Survey's year 1 data, using the self-calibration method and LSST-DESC tools}

\author[E. Pedersen et al.]{Eske M. Pedersen\orcidlink{0000-0002-8883-2172}$^{1,2,6}$\thanks{E-mail: eske.pedersen@newcastle.ac.uk},
Leonel Medina-Varela$^{2}$, Emily Phillips Longley$^{4}$, Mustapha Ishak$^{2}$, Joe Zuntz$^{3}$,
\newauthor
Chihway Chang$^{5}$, and C. Danielle Leonard$^{1}$
for the LSST Dark Energy Science Collaboration (LSST DESC)
\\
$^{1}$School of Mathematics, Statistics and Physics, Newcastle University, Herschel Building, NE1 7RU Newcastle-upon-Tyne, U.K.\\
$^{2}$Department of Physics,The University of Texas at Dallas, Richardson, TX 75080, USA\\
$^{3}$Institute for Astronomy, University of Edinburgh, Royal Observatory, Blackford Hill, Edinburgh, EH9 3HJ, U.K.\\ %ADD REST OF ADDRESS
$^{4}$Department of Physics, Duke University, Durham NC 27708, USA\\    %ADD REST OF ADDRESS
$^{5}$Department of Astronomy and Astrophysics, University of Chicago, Chicago, IL 60637, USA\\ %ADD REST OF ADDRESS
$^{6}$Department of Physics, Harvard University, 17 Oxford st., Cambridge, MA 02143, USA\\
}

\date{Accepted XXX. Received YYY; in original form ZZZ}

\pubyear{2026}

\begin{document}
\label{firstpage}
\pagerange{\pageref{firstpage}--\pageref{lastpage}}
\maketitle

\begin{abstract}
    We present the implementation of a Self-Calibration of Intrinsic Alignments of galaxies as an extension to the Vera C. Rubin Observatory's Legacy Survey of Space and Time (LSST) Dark Energy Science Collaboration (DESC)'s weak lensing 3x2pt pipeline (TXPipe). As a demonstration, we have run this pipeline on the Dark Energy Survey (DES) year one data set. We find indications of a non-zero intrinsic alignment signal in the higher redshift bins, while in the lower bins our results look more uncertain. We believe this is caused by known issues with the individual galaxies photo-z estimation. This effect is particularly harmful for the self-calibration method, since it has high requirements for reliable estimation of the photo-$z$s, and the need for individual galaxy point estimates and tomographic binning to match. We show how different methods of recreating the redshift probability distribution can affect the detection of intrinsic alignment. 
\end{abstract}

% Select between one and six entries from the list of approved keywords.
% Don't make up new ones.
\begin{keywords}
gravitational lensing: weak -- Cosmology %-- Intrinsic-alignment -- Self-calibration
\end{keywords}

%%%%%%%%%%%%%%%%%%%%%%%%%%%%%%%%%%%%%%%%%%%%%%%%%%

%%%%%%%%%%%%%%%%% BODY OF PAPER %%%%%%%%%%%%%%%%%%

\section{Introduction} %NEED TO REVISE IT

With the last decade's strong showing of results from photometric galaxy surveys like the Dark Energy Survey (DES) \citep{DES:2017gwu, DES:2017qwj}, the Hyper Suprime-Cam (HSC) survey \citep{Hamana:2019etx}, and the Kilo Degree Survey (KiDS) \citep{Hildebrandt:2016iqg}, it is clear that weak gravitational lensing is quickly becoming a central probe for investigating cosmological parameters, like $H_0, \Omega_m, S_8$, and dark energy parameters like $w_0$ and $w_a$. With upcoming stage-IV surveys such as Euclid\footnote{www.cosmos.esa.int/web/euclid, www.euclid-ec.org}\citep{laureijs2011euclid}, the Nancy Grace Roman space telescope\footnote{roman.gsfc.nasa.gov}\citep{spergel2015widefield, Dore:2019pld} and the Vera C. Rubin Observatory's Legacy Survey of Space and Time (LSST)\footnote{www.lsst.org} \citep{LSSTverarubin}, we only expect weak lensing to become even more important.

However, the weak lensing signal is not without its problems. In particular, several systematic effects seem to plague it, the two focused on here being: the photometric redshift estimation, and the intrinsic alignment of galaxies. The first shows up as issues with determining the exact distance to objects where we do not have spectroscopy, which is actively being worked on \citep{Graham_2020, Kalmbach_2020, Schmidt_2020}. The second is a more subtle systematic that causes issues at a fundamental level of weak lensing. Weak lensing is based on the idea that small but statistically significant distortions in the average shape of galaxies can be attributed to the gravitational lensing of galaxies. However galaxies are clearly not randomly distributed, but rather they are intrinsically aligned with the underlying large scale structure, this intrinsic alignment signal shows up as either a false positive or false negative in the lensed shape signal \citep{Hirata:2003cv, Heymans:2003pz, Bridle:2007ft, 2015PhR...558....1T}, which will bias any interpretation of the correlations of these shapes. If this signal is not accounted for, it is expected to create significant biases in the cosmological parameter estimation, see \citet{Mandelbaum:2005hr, Yao:2017dnt} for examples. While the signal existence is well documented, it is an actual physical signal and a significant fraction of the shape correlation budget, so it needs to be modelled very carefully. Theory development in understanding intrinsic alignments is still ongoing; see \citet{Blazek:2017wbz, Vlah:2020ovg} as examples. 

We present here an extension to the Dark Energy Science Collaboration (LSST-DESC)'s 3x2pt software pipeline (TXPipe\footnote{https://github.com/LSSTDESC/TXPipe}) \citep{TXPipe2022} that will attempt to utilize the self-calibration method suggested in \citet{Zhang:2008pw, Troxel:2014kza} to achieve a model independent estimate of the intrinsic alignment signal contaminating two-point correlations. TXPipe is designed as a modular software pipeline that will conduct the measurement part of the 3x2pt analysis for DESC, taking catalogs of galaxies positions and shapes along with needed survey information and then robustly estimating the three typical two-point correlations between galaxies; the position - position correlation (called clustering), the position - shape correlation (also known as galaxy-galaxy lensing), and shape - shape (shear - shear) which is referred to as cosmic shear. The interplay between all 3 helps in breaking some of the degeneracies that any of the probes might have \citep{DESy1-3x2pt, TXPipe2022}.

Galaxy intrinsic alignments are divided into two categories: first we have the pure intrinsic alignment -- intrinsic alignment correlation $(II)$, which arises from the fact that galaxies close to each other align toward their common neighbouring dark matter distribution. This correlation contributes as a false positive in the overall cosmic shear. However, this effect is understood to be fairly local, and a common approach is simply to avoid looking at the cosmic shear correlation within the same tomographic bin \citep{Troxel:2011za}. 
The second category includes the intrinsic alignment -- gravitational lensing correlation $(IG)$, and the related intrinsic alignment -- galaxy clustering correlation $(Ig)$. These correlations are due to the fact that the same dark matter structure that radially aligns a nearby galaxy towards it also distorts the image of a background galaxy. These correlations typically show up as anti-correlations in the signal and are not limited to the auto-correlations of bins. The usual method of only using cross-bins does not work here since these correlations are present even between remote bins. Several methods have been proposed and tried throughout the last two decades to mitigate this contamination \citep{ Troxel:2011za,Blazek:2017wbz,Blazek_2012,Leonard_2018}.
The method which we will focus on here is the self-calibration method \citep{Pedersen:2019wfp,Yao:2018pgk, Zhang:2008pw}. This method has emerged as a strong mitigation strategy to not only obtain both an isolated intrinsic alignment correlation, but also rid the cosmic shear signal of the intrinsic alignment-galaxy lensing correlation. Additionally, the self-calibration method also isolates the "clean" galaxy-galaxy lensing and galaxy-intrinsic alignment signals in source bin auto-correlations.
There has also been work on extending the method to 3-point correlations in \citet{Troxel:2011za,Troxel:2012mx,Troxel:2012pu}, along with its application to cross-correlation between cosmic microwave background lensing and galaxy intrinsic alignment \citep{Troxel:2014kza}. 
The method's clear advantage is that it does not assume any IA model to operate nor does it require any further cuts in the lensing signal data \citep{2015PhR...558....1T}. There is much to be gained from isolating the intrinsic alignment signal, for example, information about structure formation, but we focus here on obtaining the signals by themselves and leave such analysis for future work. 

We introduce our implementation of the self-calibration method into TXPipe, by applying it to the Dark Energy Survey's first year of data \citep{Hoyle_2018, Zuntz_2018}. In Section \ref{sec:Theo} we present an introduction to the underlying theory, first of two-point correlations and then of intrinsic alignment of galaxies, and the self-calibration thereof. Then, in Section \ref{sec:data_and_methods}, we discuss the Dark Energy Survey's year one data (DESy1) along with TXPipe and our extension to it. Also in this section we will briefly mention the stand-alone code needed to bridge between the photo-z estimation and the extension in TXPipe, before moving on to discussing the results in Section \ref{sec:Results}: both the results of the stand-alone estimation of the photo-z related parameters along with the extensions results, followed by their combination in isolating an intrinsic alignment signal. Finally, in Section \ref{sec:Discussion}, we will discuss the potentials of this method, along with the potential pitfalls and issues we faced with this data-set in particular.

\section{Theory} 
\label{sec:Theo}
\subsection{Two-point correlations}
Before delving into intrinsic alignment and self-calibration, we need to assess two-point correlations. We are looking at the two-point correlations of galaxy shapes, or shear, $\gamma$. Specifically, we are looking at the 2 correlations related to the shape of galaxies: cosmic shear (shape-shape) and galaxy-galaxy lensing (position-shape). Cosmic shear is often decomposed into the part tangential to the separation vector $(\vec{\theta})$ between the two galaxies and its cross component \citep{Weinberg_2013, BartelmannSchneider_2001}. The two-point correlations are therefore often written as:
\begin{align}
\label{eq:xipm}
    \xi_\pm &= \braket{\gamma_t \gamma_t}\left(\theta\right)\pm \braket{\gamma_{\times} \gamma_{\times}}\left(\theta\right)\\
    \gamma_{T/X} &= \braket{\delta_g \gamma_{t/\times}}\left(\theta\right)
\end{align}
With $\xi_\pm$ being the cosmic shear correlations, and $\gamma_{T/X}$ being the galaxy-galaxy lensing correlation. The second signal of galaxy-galaxy lensing $\gamma_X$, is expected to be a null result, and we will therefore often ignore it. When ignoring the cross part, we will refer to the galaxy-galaxy lensing signal as $w^{\gamma g}$ and we will refer to only the tangential part of Equation \ref{eq:xipm} as $w^{GG}$. 
These two-point correlations can first be related to the lensing power spectra as:
\begin{align}
    \xi_\pm &= \int_0^\infty \frac{d\ell \ell}{2\pi} J_{0/4}\left(\ell\theta\right) C_{GG}\left(\ell\right), \\
    \gamma_T &= \int_0^\infty \frac{d\ell \ell}{2\pi}J_{2}\left(\ell\theta\right) C_{\gamma g}\left(\ell\right),
\end{align}
where $C_{GG}(\ell) $ and $C_{Gg}(\ell)$ are the pure cosmic shear and galaxy-galaxy lensing power spectra respectively. The lensing power-spectra are related with underlying matter power-spectra as seen in \cite{BartelmannSchneider_2001, Weinberg_2013}. 

\subsection{Intrinsic alignment}
\label{sec:IA}
The idea of intrinsic alignment arises from the fact that when we estimate cosmic shear, we don't just obtain the gravitational lens distortion but also a signal that is related to the intrinsic alignment of each galaxy \citep{Mandelbaum:2005hr, Bridle:2007ft, 2015PhR...558....1T}. So the observed shear signal can be decomposed as:
\begin{equation}
    \gamma^\text{obs} = \gamma^G + \gamma^I + \gamma^N
\end{equation}
where $\gamma^G$ is the genuine gravitational shear we traditionally look for in weak-lensing surveys. The $\gamma^N$ is the noise part of the shear signal, while the part we are interested in is the $\gamma^I$, which is the intrinsic alignment signal. Looking at the shear-shear two-point correlation $\braket{\gamma^\text{obs}_i, \gamma^\text{obs}_j}$ between two redshift bins indicated by $i$ and $j$, we can write out the correlation as:
\begin{equation}
\label{eq:cosmicshear}
    \braket{\gamma^\text{obs}_i, \gamma^\text{obs}_j} = \braket{\gamma^G_i, \gamma^G_j} + \braket{\gamma^I_i, \gamma^G_j} + \braket{\gamma^G_i, \gamma^I_j} + \braket{\gamma^I_i, \gamma^I_j} + \text{noise}
\end{equation}
We note that the last two terms $\braket{\gamma^I_i, \gamma^I_j} + \text{noise}$ are only relevant when we are looking at the auto-correlation of a given redshift bin\footnote{Assuming uncorrelated noise between bins, and non-overlapping redshift distributions.}. The $\braket{\gamma^G_i, \gamma^G_j}$ is the actual uncontaminated cosmic shear correlation. The final two terms $\braket{\gamma^I_i, \gamma^G_j} + \braket{\gamma^G_i, \gamma^I_j}$ are the correlations between the intrinsic alignment and shear. Throughout this paper we will adopt the assumption that the first object mentioned (the $i$'th bins object) is at a lower redshift\footnote{Or at most in the same redshift bin as $j$.} than the object in bin $j$. With this, we can clearly say that the term $\braket{\gamma^G_i, \gamma^I_j}$ is going to be zero\footnote{For broad bins this might not hold. We theoretically estimated the size of the expected signals in Appendix \ref{app:signal_size}.} and we will focus on the term $\braket{\gamma^I_i, \gamma^G_j}$, which we will shorten to $IG$.

We can similarly write the two-point correlation of galaxy position and galaxy shape as:
\begin{equation}
\label{eq:galaxygalaxylensing}
    \braket{\delta^g_i, \gamma^\text{obs}_j} = \braket{\delta^g_{i}, \gamma^G_j} + \braket{\delta^g_{i}, \gamma^I_j} + \text{noise}
\end{equation}
where $\delta^g_{i}$ is the galaxy number count over-density. The galaxy over-density is related to the matter over-density usually by a galaxy bias parameter see \citet{Pandey:2020zyr}. We go into more depth on this in Appendix \ref{app:IABias}, where we also discuss the second term on right side of Equation \ref{eq:galaxygalaxylensing} in more detail. 

Currently there are 2 main methods of alleviating intrinsic alignment as a contaminant of weak-lensing measurements. The first method is modeling \citep{Blazek:2017wbz, Bridle:2007ft} where a model of the intrinsic alignment contribution is defined and parameterized. The parameters can then be incorporated in the cosmological fitting part of the analysis. Several different models have been suggested, the one used in \citet{DESy1-3x2pt} is the nonlinear-alignment model (NLA) \citep{Bridle:2007ft, Hirata:2004gc, Krause2017, Joachimi_2011}, with an expected power law scaling of the amplitude:
\begin{equation}
    A(z) = A_{IA} \left(\frac{1+z}{1+z_\text{pivot}}\right)^{\eta_{IA}}
\end{equation}
with $A_{IA}$ being an amplitude, and $\eta_{IA}$ being the power scaling with respect to redshift, while we adopt a $z_\text{pivot}=0.62$ to be consistent with \citet{DESy1-3x2pt}. The second method is self-calibration which we will cover in the next section.

\subsubsection{Self-calibration}
\label{sec:self-cal}
Here, we re-derive a version of the self-calibration method first outlined in \citet{Zhang:2008pw}. We will focus on the real space variant outlined in \citet{Pedersen:2019wfp}. 

There are two main ideas behind the self-calibration method: 
\begin{enumerate}
    \item Separating the $Ig$ signal from the $Gg$ signal when measuring galaxy-galaxy lensing by using the asymmetry of the lensing signal depending on the ordering of source and lens galaxies.
    \item Utilizing a scaling relation with the photometry information provided by a photometric survey to estimate the $IG$ term. 
\end{enumerate} 
\citet{Zhang:2008pw} originally theorized that there is a relationship between the IA signal detected in the galaxy-galaxy lensing auto-correlation and the IA-shear term between two separate tomographic bins. This concept has been explored more in other works \citep{Troxel:2014kza,Troxel:2011za} and applied to the KiDS catalog in \citet{2020MNRAS.495.3900Y,Pedersen:2019wfp}. However, before we get to the relationship between the IA-shear correlation and the IA-galaxy clustering correlation, we have to be able to separate out the terms in Equation (\ref{eq:galaxygalaxylensing}). This is done by the following argument: If we have two galaxies in the same tomographic bin, but the galaxy we measure the shear of, let us call it galaxy $S$, is behind the galaxy we have the position of, calling this galaxy $G$, then we would expect there to be lensing happening, and the shear we see would be part of the first term in Equation (\ref{eq:galaxygalaxylensing}). However if galaxy $S$ is in front of galaxy $G$, while we cannot explain the sheared signal by lensing, we can explain it with intrinsic alignment. If we knew the precise redshift of each galaxy, this would be a trivial exercise. Alas, in photometric surveys we don't have full spectra and, therefore, we don't have perfect redshift information \citep{Hoyle_2018}. This leads us to the introduction of a quality parameter $Q_i$ and a selection function $S(z_G^P, z_g^P)$. The quality parameter measures how well we are able to estimate galaxies' redshifts within the tomographic bin. Throughout this paper, we will work with 4 different redshifts: $z_g$ and $z_G$ are the "true" redshifts of the galaxies used for the position and the shear respectively, while $z_g^P$ and $z_G^P$ are the photometric redshift estimates for the same two galaxies. The selection function attempts to select galaxy pairs where the source galaxy is in front of the lens galaxy, using only photometrically determined redshifts:
\begin{equation}
    \label{eq:selection}
    S(z^P_G,z^P_g)=\begin{cases}
    1 &\text{for $z^P_G<z^P_g$}\\
    0 &\text{otherwise.}
    \end{cases}
\end{equation}
From this we can then build two observable two-point correlations. If we shift to using $w^{\gamma g}_{ii}\left(r_p\right)$ notation for the galaxy-galaxy lensing correlation in real space, then the detected signal corresponds to the left side of Equation (\ref{eq:galaxygalaxylensing}) at a separation perpendicular to line of sight $r_p$. With this we define our two observables as:
\begin{align}
    w^{\gamma g}_{ii}\left(r_p\right) &= w^{Ig}_{ii}\left(r_p\right) + w^{Gg}_{ii}\left(r_p\right)\label{eq:wgammag},\\
    \left.w^{\gamma g}_{ii}\right|_S\left(r_p\right) &= w^{Ig}_{ii}\left(r_p\right) +  \left.w^{Gg}_{ii}\right|_S\left(r_p\right)\label{eq:wgammagS}.
\end{align}
Here we have on the right hand side capital $G$ indicating the actual gravitational lensing shear, while $g$ is the galaxy number density and $I$ is the intrinsic alignment shear. 
Since the intrinsic alignment signal should not depend on the ordering within the bin\footnote{Assuming that the redshift bins are thin enough that there's no significant IA redshift evolution throughout the bin.}, we expect their contribution to be the same in both Equation (\ref{eq:wgammag}) and Equation (\ref{eq:wgammagS}). Meanwhile, we expect a smaller signal from the selection function correlation than for the non-selection function case: $\left(\left.w^{Gg}_{ii}\right|_S \left(r_p\right)< w^{Gg}_{ii}\left(r_p\right)\right)$, since the selection function tries to pick out the part of the correlation where there shouldn't be a signal. If we had perfect positional knowledge we would expect $\left.w^{Gg}_{ii}\right|_S  = 0$. From this we define a photo-$z$ "quality ratio" $Q_i$ as:
\begin{equation}
\label{eq:Q}
    Q_i\left(r_p\right) = \frac{\left.w^{Gg}_{ii}\right|_S\left(r_p\right)}{w^{Gg}_{ii}\left(r_p\right)}
\end{equation}
Having $Q_i=0$ indicates a perfect photo-z quality, that is we can predict the redshift position of all galaxies as if they were spectroscopically measured. $Q_i=1$ indicates a complete breakdown in the estimation of photo-z´s such that any attempt at determining the internal distribution within a given redshift bin is impossible. 
We can theoretically describe the two correlations in Equation (\ref{eq:Q}) as functions of the angular separation $\theta$ and then relate $\theta$ and $r_p$.

\begin{align}
    w^{Gg}_{ii}\left(\theta\right) &= \frac{1}{2\pi}\int_0^\infty \frac{W_i\left(\chi\right)n_i\left(\chi\right)}{\chi} \times \nonumber \\
    &\quad \int \ell b_g P_\delta \left(k=\frac{\ell}{\chi}; \chi\right)J_2\left(\ell\theta\right)d\ell d\chi \label{eq:wGg}\\
    \left. w^{Gg}_{ii}\right|_S \left(\theta \right) &= \frac{1}{2\pi}\int_0^\infty \frac{W_i\left(\chi\right)n_i\left(\chi\right)}{\chi}\eta_i\left(z\right) \times\nonumber \\
    &\quad \int \ell b_g P_\delta \left(k=\frac{\ell}{\chi}; \chi\right)J_2\left(\ell\theta\right)d\ell d\chi \label{eq:wIg}
\end{align}

In Equation (\ref{eq:wGg}) and (\ref{eq:wIg}): $W_i$ is the lensing kernel; $n_i$ is the true galaxy distribution; $\chi$ is the comoving distance; $J_2$ is the second order Bessel function (as we are conducting a Hankel transformation) and $P_\delta$ is the power spectrum. Finally there is $\eta_i$ which is the function given in \cite{Zhang:2008pw} as shown in Equation (\ref{eq:eta}):

\newpage
\begin{widetext}
\begin{equation}
\eta_i(z) =\frac 
	{2\int_{z^P_{ i, \rm min}}^{z^P_{ i, \rm max}}\mathrm{d}z^P_{G}\int_{z^P_{ i, \rm min}}^{z^P_{ i, \rm max}}\mathrm{d}z^P_g\int_{0}^{\infty}\mathrm{d}z_G W_L(z,z_G)p(z_G|z^P_G)p(z|z^P_g)S(z^P_G,z^P_g)n^P_i(z^P_G)n^P_i(z^P_g)}
	{\int_{z^P_{ i, \rm min}}^{z^P_{ i, \rm max}}\mathrm{d}z^P_{G}\int_{z^P_{ i, \rm min}}^{z^P_{ i, \rm max}}\mathrm{d}z^P_g\int_{0}^{\infty}\mathrm{d}z_G W_L(z,z_G)p(z_G|z^P_G)p(z|z^P_g)n^P_i(z^P_G)n^P_i(z^P_g)}  \label{eq:eta}
\end{equation}
\end{widetext}

Here the superscript $P$ indicates a photometrically estimated redshift and $n_i^P$ are the corresponding distributions. $p\left(z|z^P\right)$ is the photometric distribution function or PDF for objects in our survey, while $W_L$ is the window function which is given as:
\begin{eqnarray}
   W_L\left(z_L, z_S\right) = \begin{cases}
   \frac{3}{2}\Omega_m \frac{H_0^2}{c^2}\left(1+z_L\right) \chi_L \left(1-\frac{\chi_L}{\chi_S}\right) &\text{for}\, z_L < z_S\\
   0 &\text{otherwise.}
   \end{cases}
\end{eqnarray}
$S$ is the selection function given in Equation (\ref{eq:selection}). The integration of $z_G^P$ and $z_g^P$ in Equation (\ref{eq:eta}) are over the span of the $i$'th redshift bin.

With the Quality parameter we next split the two Equations \ref{eq:wgammag} and \ref{eq:wgammagS} into their physical parts, rather than observational:
\begin{align}
    w^{Gg}_{ii} &= \frac{w^{\gamma g}_{ii}\left(r_p\right) - \left.w^{\gamma g}_{ii}\right|_S\left(r_p\right)}{1-Q_i} \label{eq:wGg2}\\
    w^{Ig}_{ii} &= \frac{\left.w^{\gamma g}_{ii}\right|_S\left(r_p\right) - Q_i\times w^{\gamma g}_{ii}\left(r_p\right)}{1-Q_i} \label{eq:wIg2}  
\end{align}

The separation above assumes that the density field of the source sample traces galaxy clustering alone. However, weak lensing by foreground structure also modulates the observed number density of the galaxies used as density tracers, a feature known as magnification bias \citep{BartelmannSchneider_2001}. Like the IA term, this contribution does not depend on the relative ordering of the pair along the line of sight, so Equations (\ref{eq:wGg2}) and (\ref{eq:wIg2}) return $w^{Gg}_{ii}$ essentially unchanged, but this does affect $w^{Ig}_{ii}$ \citep{Yao2023}: magnification contaminates the separated intrinsic-alignment signal specifically. We do not model this term in the present analysis and leave its treatment to future work; we discuss its implications in Appendix~\ref{app:magnification}.

The final step is to estimate the connection between $w^{Ig}_{ii}$ and $w^{IG}_{ij}$ as in \citet{Pedersen:2019wfp}. Under the assumption of linear galaxy bias, we introduced the relationship in the real space to be:
\begin{equation}
\label{eq:scalingrelation}
    w^{IG}_{ij}\left(r_p\right) \approx \frac{W_{ij}\Delta_i}{b_i}w_{ii}^{Ig}\left(r_p\right)
\end{equation}
where $W_{ij}$ is the weighted lensing kernel between the two bins:
\begin{equation}
    W_{ij} = \int_0^\infty \mathrm{d}z_L \int_0^\infty \mathrm{d}z_S W_L\left(z_L,z_S\right) n_i\left(z_L\right)n_j\left(z_S\right),
\end{equation}
and $\Delta_i$ is the effective bin width:
\begin{equation}
    \Delta_i^{-1} = \int_0^\infty n_i^2\left(z\right) \frac{\mathrm{d}z}{\mathrm{d}\chi}\mathrm{d}z,
\end{equation}
The resulting $W^{IG}_{ij}$ is in fact a linear combination of intrinsic alignment term measured in $\xi_+$ and $\xi_-$.
We note that in deriving Equation (\ref{eq:scalingrelation}) we assumed a linear galaxy bias $b_i$, so that we can relate the intrinsic alignment-matter correlation with the intrinsic alignment-positioning correlation:
\begin{equation}
    \frac{w_{ii}^{Ig}\left(r_p\right)}{b_i\left(r_p\right)} = w_{ii}^{Im}\left(r_p\right)
\end{equation} 

We show in Appendix \ref{app:IABias} that this holds for at least second order galaxy bias.

One thing to note here is that the Dark Energy Survey distinguishes between lenses and sources for galaxy-galaxy lensing. For our analysis we will only be looking at the source part of the catalog, since the cosmic-shear signal is calculated only from the sources. This means that even in our derivation of $\gamma_{ii}^{Ig}$ the $\delta_g$ is obtained from the source catalog's $i$'th bin. 

\section{Data and method}
\label{sec:data_and_methods}
\subsection{Dark Energy Survey year 1}
We make use of the Dark Energy Survey year 1 data release (DES year 1), which is publicly available. We specifically use the Metacalibration (Metacal) shape catalog \citep{Zuntz_2018} and the photo-$z$ catalogs.
The Metacal shape catalog is described in details in \citet{Zuntz_2018}, while the DES y1 photometric redshift estimation is described in \citet{Hoyle_2018}. A key thing to note is that the focus of this photo-$z$ estimation is not the individual galaxy but rather the assigning of galaxies to a redshift bin and the estimation of a redshift distribution for such a bin. 
We use the catalogs in two instances: we use it to generate the photometric redshift distribution as described in Section \ref{sec:PRD}; and we use the shape catalogs and photo-z catalogs to do a similar data reduction to the original DESY1 3x2pt analysis \citep{DESy1-3x2pt}, as outlined in \citet{LSSTDarkEnergyScience:2022amt}. The reduced data is fed into DESC's 3x2pt pipeline TXPipe to obtain the needed correlations. The DES year 1 catalog consists of 4 tomographic redshift bins, with numbering 1 through 4. 

\subsection{TXPipe}
\label{sec:TXPipe}
TXPipe is the 3x2 point correlation software that is being built within DESC for the upcoming Legacy Survey of Space and Time (LSST). This is a complete software package that takes inputs like the raw Metacal catalogs and photometry information. From this, TXPipe takes care of the actual calibration, selection and cuts in the data, including doing consistency checks and calculating maps and harmonic counterparts to the real-space data. For more information on this see \cite{TXPipe2022, LSSTDarkEnergyScience:2022amt}. We use the extension for TXPipe that calculates the two correlations we need for the self-calibration in the form of Equations \ref{eq:wgammag} and \ref{eq:wgammagS}. TXPipe uses several well-established Python modules such as TreeCorr\footnote{github.com/rmjarvis/TreeCorr} and PyCCL\footnote{github.com/LSSTDESC/CCL} \citep{Chisari:2018vrw}. 
The extension includes a new subclass of TXpipes "two-point correlation" class, that includes galaxy-galaxy lensing between two source catalogs. It also converts inputs to TreeCorr \citep{Jarvis:2003wq} so we can convert the outputs of TreeCorr from $r_p$ to $\theta$. This is done by defining a fiducial cosmology which can be used to estimate the initial radial distances for each bin from the mean redshift of the bin. 
We then use PyCCL \citep{Chisari:2018vrw} to convert the redshift of each galaxy to a co-moving distance and their separations ($r_p$) in Mpc to $\theta$, using the mean redshift of the bin as the radial distance in the conversion. We define separation bins for each redshift bin in Mpc, such that they correspond to separations in $\theta$. For this project we use the fiducial cosmology given in \cite{DESy1-3x2pt}. 

\subsection{SCIA2020}
\label{sec:SCIA2020}
The calculations presented in Section \ref{sec:self-cal} that are not related to the correlation functions are contained in a series of Jupyter notebooks\footnote{https://jupyter.org}  and python scripts, we call "Self-Calibration of Intrinsic Alignment 2020" (SCIA2020)\footnote{github.com/empEvil/SCIA2020}. This includes the code used to estimate $\eta$ as given in Equation \ref{eq:eta} and generating the PDFs for each bin. Many of the integrals for obtaining $\eta$ have been calculated by Monte-Carlo integration. For future development on this part of the code see the discussion section. This also includes the use of PyCCL \citep{Chisari:2018vrw} to attempt to obtain the $Q(\theta)$. 

To verify our implementation with TXpipe and SCIA2020 we constructed a second independent pipeline outlined in Appendix \ref{app:Secondpipe}.
  
\section{Results}
\label{sec:Results}
\subsection{Photometric redshift distribution}
\label{sec:PRD}
Looking at the photometric redshifts, and the distributions made available for DES year 1 \citep{Hoyle_2018} we see that focus is largely on ensuring understanding of and stable tomographic redshift bins. As they noted in \citet{Hoyle_2018}, their weak-lensing results are not sensitive to the details of the distribution of galaxies within bins, as long as the tomographic bins have well estimated redshift means, that are well separated from each other. This means that we have several considerations in our recreation of the photo-$z$'s. We are also limited by the data available on a per galaxy basis. We therefore cannot recreate the probability distribution functions (PDFs) for the individual galaxies. In the end we have several different options:
\begin{enumerate}
    \item We recreate the PDF by using the $z_\text{mean}$ (which is the mean photometric estimated redshift) as the photo-$z$ for each galaxy while using the $z_{mc}$ as "true" redshift. This will, on the overall distribution, recreate the redshift distributions that DES year 1 got in \citet{Hoyle_2018}. This also requires that we do a weighting of each galaxy depending on its MetaCal response and selection function. We will call this our weighted PDF.
    \item We could do the same but leave out the weighting, which slightly changes the distribution. We will refer to this as raw PDF. 
    \item We could, instead of trying to recreate the redshift distributions, look at the information we have of each galaxy, i.e. we know its mean redshift ($z_\text{mean}$) and an estimated error ($z_\sigma$). Using these two values we could estimate a Gaussian probability distribution for it, and then stack up these to come up with our 2-dimensional PDF: 
    \begin{equation}
\label{eq:Gaussianpz}
    p(z, \bar{z}) = \frac{1}{\sqrt{2\pi \sigma}}\exp \left( -\frac{1}{2} \left(\frac{z - \bar{z}}{\sigma}\right)^2 \right)
\end{equation}
    \item Finally we could do the same as above but instead of centering our Gaussian probabilities on the $z_\text{mean}$ we could center it on the peak of the probability distribution ($z_\text{mode}$). Particularly for the first and last tomographic redshift bins, these numbers don't coincide with each other. 
    \item we could do the two suggested Gaussian methods, while using the MetaCal weights as well. 
\end{enumerate}
We should note that as pointed out in the Appendix of \citet{Schmidt_2020}, there is quite a bit of subtlety in reducing PDFs down to a point estimate. With the mean often ending up in lower probability areas of the range, while the mode (the highest probability point) ignoring the posterior of the PDF.

\begin{figure}
    \centering
    \includegraphics[width=\columnwidth]{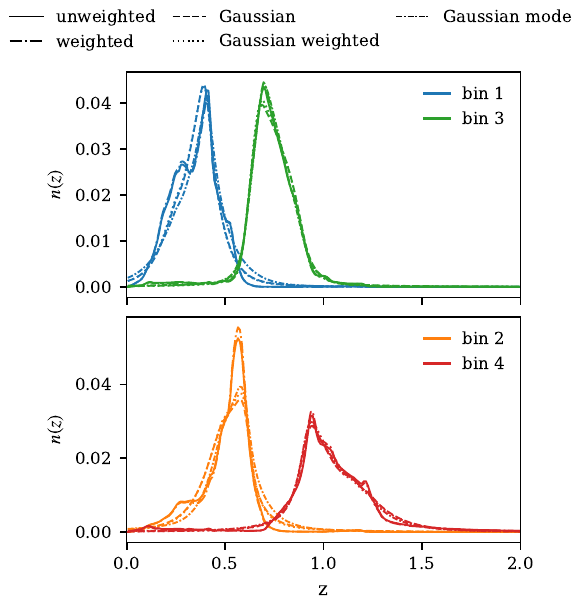}
    \caption{Examples of how the "True" redshift distribution changes, depending on the choices of how we recreate the PDF. the solid lines are the unweighted recreation using $z_\text{mean}$ and $z_{mc}$, while the dash-dotted line is the same but weighed by the MetaCal response, these lines should correspond to the original $n(z)$ used by DES. The dashed lines are the Gaussian stacked method, and the dotted line is the Gaussian stacked method, but each galaxy weighted by the MetaCal response. The lines colored purple, marron, pink, and grey are the Gaussian estimates, but centered on the $z_\text{mode}$ rather than the means.}
    \label{fig:redshift_dists}
\end{figure}

As we can see from Figure \ref{fig:redshift_dists}, there are differences even on the ensemble level depending on our method of recreating the PDF. This also means that different methods can result in different values of $\eta$, and in turn, the quality parameter. The redshift distributions here seem in decent agreement with the ones used in the analysis of DES-Y1\footnote{the weighted lines in Figure \ref{fig:redshift_dists} should be the same as the DES-Y1 analysis.} \citep{DES:2017gwu, DES:2017qwj, DESy1-3x2pt, LSSTDarkEnergyScience:2022amt}, but as we pointed out, these are for the whole of the tomographic bin, and not for individual galaxies, they therefore might not match at the individual galaxy level \citep{Hoyle_2018}.

\subsection{$\eta$ and the photometric quality parameter}
As mentioned in the previous section, we are limited by how accurately we can recreate the PDFs, and this will affect how accurate the $\eta_i$ is. $\eta_i$ (Equation \ref{eq:eta}) is, to first order, equal to the quality parameter $Q_i\left(\theta\right)$ (\ref{eq:Q}) according to \citet{Zhang:2008pw}, and will be needed if we want to obtain an estimate of $Q_i$. The advantage of using the first order approximation $Q_i \approx \eta_i$ would be that $\eta$ have very little cosmology dependence\footnote{Only the lensing kernel is cosmology dependent, and is essentially canceled out.}. For $Q_i$ we do have to define a cosmology, and we found that a variation of $\pm 20\%$ in both $h$ and $\sigma_8$ results in a about $\pm2\%$ variation in $Q$ (see Appendix \ref{app:Qcosmo}). We calculate $\eta$ using Equation \ref{eq:eta}, and using SCIA2020 as described in Section \ref{sec:SCIA2020}.

\begin{figure*}
    \centering
    \includegraphics{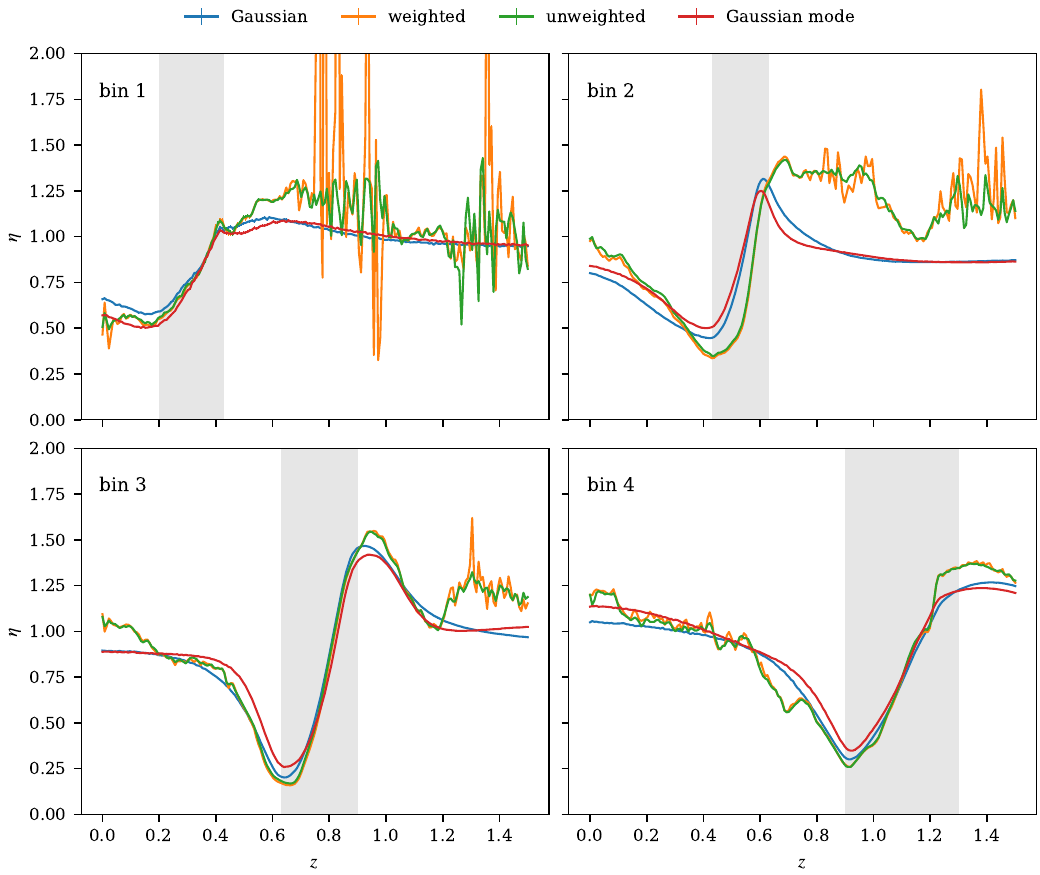}
    \caption{We illustrate here the values of $\eta$ for each of our four bins.  We used a Monte-Carlo integrator, and the errors reflects this. We estimated $\eta$ using 4 different PDFs. The blue is the Gaussianized method, the orange is the Metacal weighted one, the green is the unweighted PDF, and the red is using the Gaussianized method centered on the $z_\text{mode}$ . The grey band highlights the redshift span of the tomographic bin that the $\eta$ corresponds to.}
    \label{fig:eta_bins}
\end{figure*}

First, lets compare the $\eta(z)$ obtained from the different PDFs. These are shown in Figure \ref{fig:eta_bins}. For all bins and realizations, we used the same Monte-Carlo integration mentioned in Section \ref{sec:SCIA2020}. This also means that we used the same number of points for the integration. For the results presented here we did 200 steps in redshift from 0 to 1.5, and for each step we did 1 million points in the integration. The errors shown in Figure \ref{fig:eta_bins} are estimated from the integration. 
It can be seen that for all but the first bin the integration seems to have very little uncertainty, interestingly it is all four realizations that seem to have integration uncertainty in them. We also note that for the Non-Gaussian realizations outside the bin areas, there seems to be some instability in the results. This is especially clear in at the high redshift end, and in bin 3 and 4 at low redshifts. 

For bins 3 and 4, for large parts of the range of their relevant bins, the 4 realizations give the same values for $\eta$ with the caveat that at the higher end of bin 4 there seems to be a bit of a discrepancy with the "Non-Gaussian" PDFs. 
More interestingly, in bin 2 we find that there is a disagreement between the Gaussian methods and the methods relying on the recreated PDFs ($z_{mc}$ or Non-Gaussian generated). We see that the Gaussian $\eta$ does not go as low in the beginning and it peaks just around the edge of the bin, rather than after bin. This also means that the slope of the two methods do not agree.

\begin{table*}
    \centering
    \caption{Here are our 6 different $\eta$ estimators. First we have the unweighted (raw) photo-z's with a flat weighted mean of the area. Next we have $\eta$ estimated from the Gaussianized distribution, but still without weighting the average. Next we have the same two distributions using respectively the photometric redshift distribution $\left(n_i^p\right)$ and the "true" redshift distributions $\left(n_i\right)$ as weights, in the latter case we of course used the true redshift distribution corresponding to the PDF used to calculate the $\eta$. The error included is estimated from the variance around the mean.}
    \begin{tabular}{cccccccccc}
        \hline
         bin & $\bar{\eta}_\text{raw,flat}$ & $\bar{\eta}_\text{Gaussian,flat}$ & $\bar{\eta}_\text{mode,flat}$ & $\bar{\eta}_{\text{raw}, n^p_i}$  & $\bar{\eta}_{\text{Gaussian}, n^p_i}$ & $\bar{\eta}_{\text{mode}, n_i^p}$ & $\bar{\eta}_{\text{raw}, n_i^r}$ & $\bar{\eta}_{\text{Gaussian}, n_i^g}$  & $\bar{\eta}_{\text{mode}, n_i^m}$  \\ 
\hline
bin 1: & $0.76\pm0.14$ &  $0.78\pm0.13$ &  $0.73\pm0.15$ & $0.87\pm 0.18$ &  $0.88\pm0.17$ &  $0.85\pm0.19$ & $0.85\pm0.17$ &  $0.87\pm0.16$ & $0.85\pm0.19$ \\ 
bin 2: & $0.59\pm0.26$ &  $0.8\pm0.29$ &  $0.85\pm0.25$ & $0.58\pm 0.26$ &  $0.79\pm0.29$ &  $0.84\pm0.25$ & $0.72\pm0.29$ &  $0.83\pm0.29$ & $0.89\pm0.26$ \\ 
bin 3: & $0.61\pm0.41$ &  $0.65\pm0.4$ &  $0.61\pm0.34$ & $0.55\pm 0.41$ &  $0.59\pm0.4$ &  $0.57\pm0.34$ & $0.59\pm0.41$ &  $0.64\pm0.4$ & $0.6\pm0.34$ \\ 
bin 4: & $0.74\pm0.36$ &  $0.71\pm0.31$ &  $0.75\pm0.3$ & $0.61\pm 0.38$ &  $0.6\pm0.32$ &  $0.64\pm0.32$ & $0.65\pm0.37$ &  $0.65\pm0.31$ & $0.68\pm0.31$ \\ 
        \hline
    \end{tabular}
    \label{tab:eta_estimators}
\end{table*}

Next we will look at $\eta_i$ as the first order prediction of $Q_i$, which means we want to obtain a point value for it across our tomographic bins. As clearly illustrated in Figure \ref{fig:eta_bins}, $\eta_i$ changes across its bin, and hence if we wish for one value of $\eta$ to encapsulate the first order approximation of $Q$ for each bin, there is not a single good answer. Here we have chosen the mean values in each bin, and then we have varied what kind of weighting scheme we have used to obtain the mean. In Table \ref{tab:eta_estimators} we have outlined the results of this, we have only looked at the three unweighted cases, since Figure \ref{fig:eta_bins} shows that within each bin there is very little difference between the unweighted and weighted version. We have estimated the mean using a flat weight across the tomographic bin $\left(\eta_\text{flat}\right)$ in the three first columns, as we can see, there is a large discrepancy in the value of $\eta_i$ for the second bin. The fourth through sixth columns are using the photometric redshift distribution as weights, and show a similar discrepancy. Finally, the last three columns show us the means if we were to use the respective "true" redshift distributions as the weights. Again we see a discrepancy in the value for bin 2 that is much larger than for any of the other bins, though still within quoted errors. 
If we look at the errors quoted in Table \ref{tab:eta_estimators}, these are the errors from comparing the mean to all the results within the bin, i.e. they are not the uncertainty, but rather a measure of the change across the bin compared to the mean value. An example is that in bin 3 we see a very large difference between the beginning and the end of the bin and this is reflected in the $\pm0.4$ that is given for all those means.
Again, this supports the the point that we have to be very careful when we wish to replace the singular values $Q_i$ with something we obtain from $\eta_i$. 

\begin{figure}
    \centering
    \includegraphics{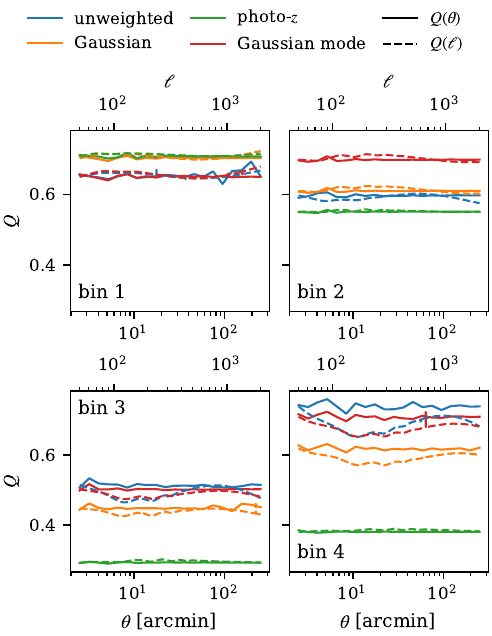} 
    \caption{We here show each of the 4 bins and the results of the different $Q$'s in each bin. The blue lines are the $Q$'s for the "raw" redshift distribution. the Orange lines are the "Gaussian" distribution. The green lines are using the "Gaussian" $\eta_i$ and then assuming we have a $n_i^p$ as our redshift distribution. The full lines are the $Q\left(\theta\right)$, while the dashed lines are $Q\left(\ell\right)$.}
    \label{fig:Q_of}
\end{figure}

Finally we can also estimate $Q\left(\theta\right)$ as expressed in Equation \ref{eq:Q} by using PyCCL \citep{Chisari:2018vrw}. We did this for all 4 bins using $50<\ell<2000$ for the power-spectra and the best-fit cosmology given by DES \citep{DESy1-3x2pt}. We did this using the 2 different $\eta_i$s we listed the means for in Table \ref{tab:eta_estimators}. We also did a third attempt where we used the Gaussian $\eta_i$ but instead used the photometric redshift distribution for the redshift distribution that goes into the CCL tracers. 
As seen in Figure \ref{fig:Q_of} we find that for the 2 higher redshift bins, the photometric $Q$ seems to be performing better than the Gaussian and raw $Q$'s, while for the 2 lower redshift bins, they perform about the same. It is also interesting to note that for the higher bins, there is a much larger spread in the results, compared to the lower redshift bins. At the same time when we consider these in comparison with the $\bar{\eta}$ we found previously, the $Q$'s are almost always lower, the exception being the 4th bin. As a reminder, all this is done using redshift distributions, not on the individual galaxy level, but rather on a ensemble level. That is, we have not looked at the individual galaxies redshift, for estimating the $\eta$ and the $Q_i$. 

\subsection{Correlation calculations}
\begin{figure}
    \centering
    \includegraphics{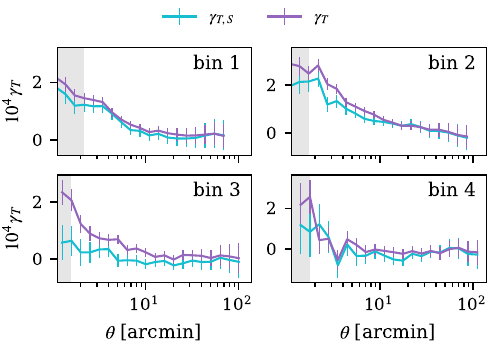} %REMAKE GRAPH WITH BIN LABELS ON TOP
    \caption{The resultant $\gamma_T$ (purple) and $\gamma_{TS}$ (cyan) corresponding to the left-hand side of Equations \ref{eq:wgammag} and \ref{eq:wgammagS} for all 4 bins. The error bars are the jackknife-estimated errors from TreeCorr (100 Jackknife regions were used). Each bin is calculated with separation $(\theta$ along the x-axis so it can be compared to the angular separations used for cosmic shear in \citep{LSSTDarkEnergyScience:2022amt}. The greyed out areas are the scale cuts suggested by \citet{DESy1-3x2pt, LSSTDarkEnergyScience:2022amt}.}
    \label{fig:gamT}
\end{figure}

For correlation calculations we used the software package TXPipe, which is being designed by LSST-DESC for the upcoming LSST survey \citep{TXPipe2022}. Our extension to TXPipe is described in more detail in Section \ref{sec:TXPipe}. For each of the 4 bins we estimate the correlations given by Equation \ref{eq:wgammag} and Equation \ref{eq:wgammagS} (also referred to as $\gamma_T$ and $\gamma_{TS}$ respectively) as seen in Figure \ref{fig:gamT}. As part of the estimation we do a scale cut, and exclude the lowest separation points from the analysis (see the shaded out area in Figure \ref{fig:gamT}). This is done under suggestion from \cite{DESy1-3x2pt,LSSTDarkEnergyScience:2022amt}, see these for more detail. This also means we will exclude these data points from any estimation of detection. 

\begin{figure}
    \centering
    \includegraphics{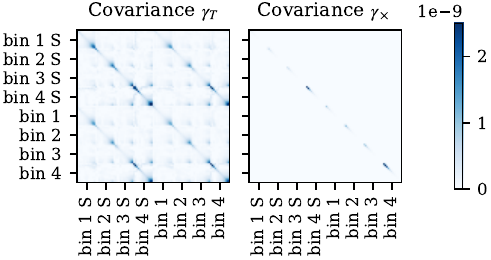}
    \caption{(Left) The jackknife-estimated covariance and cross covariance corresponding to the correlations $\gamma_T$ and $\gamma_{TS}$. (Right) The shot-noise covariance for the $\gamma_\times$ and $\gamma_{\times S}$ terms. The upper left quadrants of each graph are the covariances for the $\gamma_{Ts / \times s}$ signals, while the lower right quadrants are the covariances for the $\gamma_{T / \times}$ signal.}
    \label{fig:covariance}
\end{figure}

\begin{figure}
    \centering
    \includegraphics{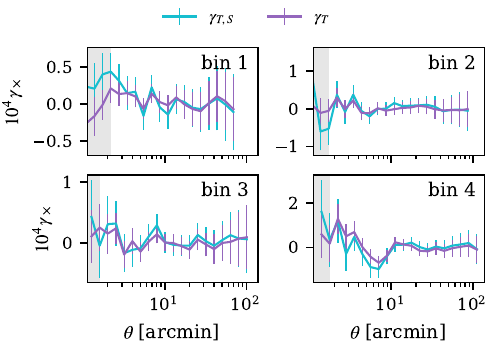} % REMAKE GRAPH WITH BIN LABELS ON TOP
    \caption{The $\gamma_\times$ (purple) and $\gamma_{\times S}$ (cyan) signals for the 4 bins. The errorbars are the shot-noise error coming from TreeCorr. The separation along the X-axis is the same separations as in Figure (\ref{fig:gamT}). Similarly the greyed out areas are the areas excluded by scale cuts.}
    \label{fig:gamX}
\end{figure}

We use the built-in TreeCorr code to estimate jackknife covariance for all these correlations, as shown in Figure \ref{fig:covariance}. The covariance is propagated to the end result in $w_{ii}^{Gg}$ and $w_{ii}^{Ig}$. We also calculated the $\gamma_\times$ and $\gamma_{\times S}$ signal as a null test, seen in Figure \ref{fig:gamX}. For the $\gamma_\times$ errors we used the shot noise error estimated by TreeCorr. As discussed later, we estimated all these correlations without relying on random catalogs. 

\subsection{Intrinsic alignment galaxy correlations}
\label{sec:Ig}
Next to separate out the $w^{Ig}_{ii}$ and $w^{Gg}_{ii}$ for each of the 4 source bins from the correlations seen in Figure \ref{fig:gamT}, we need to consider what $\eta$ or $Q's$ we can rely on. We have chosen here to present the results based on the Gaussian estimated $\eta$ and $Q$, see Table \ref{tab:eta_estimators} and Figure \ref{fig:Q_of} for details. We chose this to illustrate the limits in the information of the photo-z data. The Gaussian version takes into account the estimated uncertainty that is in the individual galaxy estimates, which isn't as well captured in the recreation attempt.

\begin{figure*}
    \centering
    \includegraphics{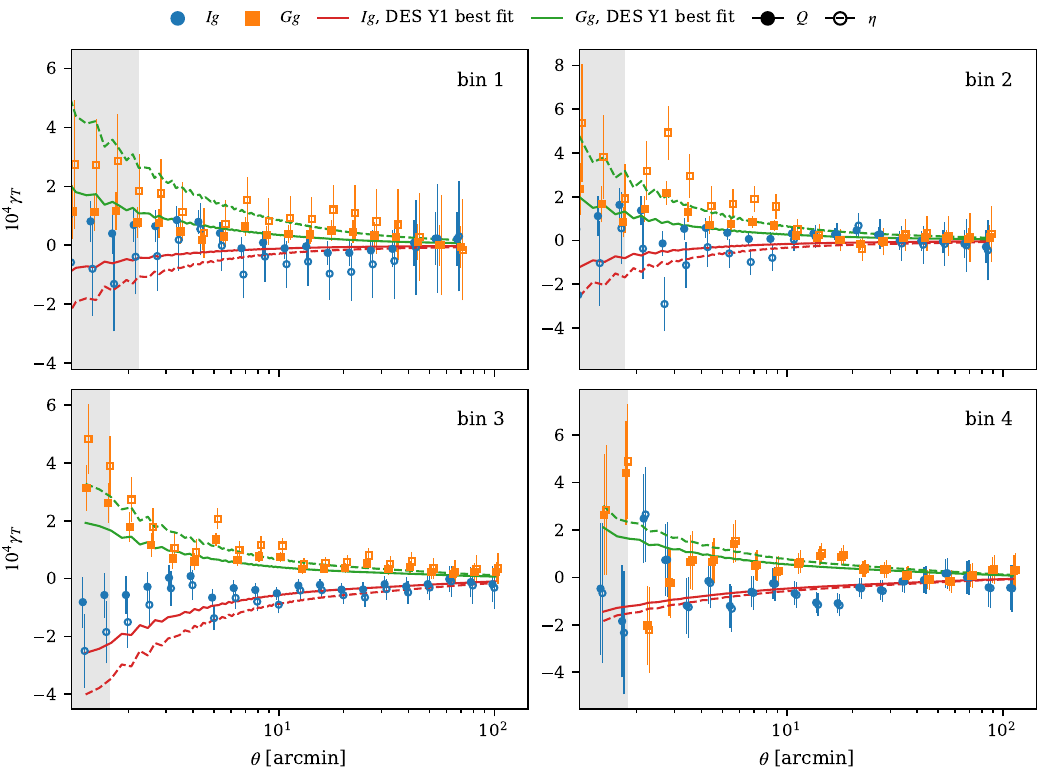}
    \caption{The separated correlations for galaxy-galaxy lensing between two source bins in DESy1. The data-points plotted are the "cleaned" galaxy galaxy lensing correlation $(Gg)$ (orange) and the intrinsic alignment-galaxy clustering correlation $(Ig)$ (blue), the dotted versions are using  the Gaussian $\eta$ from Table \ref{tab:eta_estimators} to separate the signals, while the full lines are using the Gaussian $Q$ to separate. Also plotted are the theoretical predictions for each of these correlations using the best-fit parameters given by the DES year 1 analysis, $Gg_\text{theo}$ (green) and $Ig_\text{theo}$ (red), here again because the galaxy bias is estimated from the $Gg$ correlation, the dashed lines represent the theoretical prediction with $b_i$ estimated using the $Gg_\eta$ correlations. The gray area indicates the cuts suggested in \citet{LSSTDarkEnergyScience:2022amt}.}
    \label{fig:IgGg}
\end{figure*}

Using this we obtain the correlations seen in Figure \ref{fig:IgGg}. The theoretical lines shown rely on a galaxy bias$(b_i)$ estimated from the $w^{Gg}_{ii}$ as such\footnote{This result follows from a similar argument to the one laid out for intrinsic alignment in Appendix A.}:
\begin{equation}
    \label{eq:bias_Gg}
    \tilde{b}_i(r_p)= \frac{w_{ii}^{Gg}(r_p)}{w^{Gm}_{ii, \text{Theo}}(r_p)},    
\end{equation}
where the theoretical prediction correlation comes from CCL \citep{Chisari:2018vrw}. We have added two lines because the $w^{Gg}_{ii}$ depends on if it was estimated with $\eta_i$ or $Q_i$. To estimate how well we separate out the signals, we have compared the intrinsic alignment signal to a model where it is 0, the results of this are summarized in Table \ref{tab:Ig0}, we will later also compare to the intrinsic alignment model outlined in Section \ref{sec:IA}. To estimate $\sigma$ we included a "Hartlap" correction to the precision matrix formed from the Jackknife covariance shown in Figure \ref{fig:covariance} as suggested in \citet{Hartlap2007, Percival2022MNRAS}. We include a correction factor each time we do this. 

\begin{table}
    \centering
    \caption{Comparing the $w^{Ig}_{ii}$ signal to 0. For each of the 4 bins we have estimated the number of $\sigma$ that the result differ from zero. This is done both relying on $Q$ and on $\eta$ as the quality parameter in Equation \ref{eq:wIg2}. Here $p$ and $p_\eta$ are the p-values associated with the two estimates.}
    \label{tab:Ig0}
    \begin{tabular}{ccccc}
    \hline
         bin & $\sigma$ & $\sigma_\eta$ & $P$ & $P_\eta$ \\
         \hline
1 & 0.116 & 0.00175 & 0.0923 & 0.0014 \\ 
2 & 0.105 & 0.0264 & 0.0839 & 0.0211 \\ 
3 & 0.216 & 1.15 & 0.171 & 0.749 \\ 
4 & 0.562 & 0.523 & 0.426 & 0.399 \\ 
\hline
    \end{tabular}
\end{table}

We see only a slight signal in the higher two of the redshift bins, However the signal is weak and is only in contention with a null result in the 3rd bin if we are using the $\eta$ estimated results, and then only at a $1.86\sigma$. 
We caution that bins 3 and 4 are the bins in which cosmic magnification, which the separation attributes to $w^{Ig}_{ii}$ (Section~\ref{sec:self-cal}), is expected to be largest. The separated signal should therefore be read as $w^{Ig}_{ii}+g_\text{mag}w^{G\kappa}_{ii}$, and depending on the sign of $g_\text{mag}$ for the source sample, magnification can either mimic or partially cancel an intrinsic-alignment signal (Appendix~\ref{app:magnification}). The NLA comparisons below do not include this term.

We could also consider how well the $Ig$ correlations fit with the theoretical signal. We compare here to the nonlinear-alignment model (NLA) as outlined in Section \ref{sec:IA}. We compare to the model parameters obtained in \citet{DESy1-3x2pt} and we find for the third bin a reduced $\chi^2$ of $2.52(2.02)$ for using $Q(\eta)$ respectively, and $0.731(0.736)$ for the fourth bin. We could also take a look at how the $Gg$ correlations corresponds to the theoretical estimate for it, here we find a reduced $\chi^2$ of $1.04 (0.879)$ using $Q (\eta)$ for the third bin and $0.664 (0.884)$ for the fourth bin. Lastly, we also compared the two separated signals. To do this, we calculated the Mahalanobis distance between the two sets of correlations, we summarized the results of this in Table \ref{tab:MD_Gg_Ig}. The Mahalanobis distance is a statistical tool used to generalize the distance between two data vectors with respect to an underlying probability distribution \citep{Mahalanobis}\footnote{or see: \url{https://en.wikipedia.org/wiki/Mahalanobis_distance}}:
\begin{equation}
    d_M\left(\vec{x},\vec{y}\right) = \sqrt{\left(\vec{x}-\vec{y}\right)^T \left(h\times C^{-1}\right) \left(\vec{x}-\vec{y}\right)}
\end{equation}
Here $d_M$ is the Mahalanobis distance, $\vec{x}$ and $\vec{y}$ are the two data vectors, while $C^{-1}$ is the inverse of the covariance matrix, and $h$ is the Hartlap factor. 

\begin{table}
    \centering
    \caption{The Mahalanbois distance between the $w^{Gg}$ and $w^{Ig}$ correlations for the 4 bins. Listed is the value for both the $Q$ calculated version and for the $\eta$ calculated versions. Note that these are calculated without using the cuts defined earlier.}
    \label{tab:MD_Gg_Ig}
    \begin{tabular}{ccc}
    \hline
bin & $\Delta \sigma_{Q}$ & $\Delta \sigma_\eta$ \\
\hline
 1 & 3.03 & 8.72 \\ 
 2 & 3.6 & 12.1 \\ 
 3 & 5.12 & 8.68 \\ 
 4 & 3.67 & 3.98 \\ 
 \hline
    \end{tabular}
    
\end{table}

As we can see in table (\ref{tab:MD_Gg_Ig}) only the 3rd bin has a separation between the two signals ($Ig$ and $Gg$) that is above 5 $\sigma$, with a 5.12$\sigma$ in the $Q$ version, and a $8.68\sigma$ for the $\eta$ version. 

\subsection{Intrinsic alignment in the cosmic shear}
\begin{figure}
    \centering
    \includegraphics{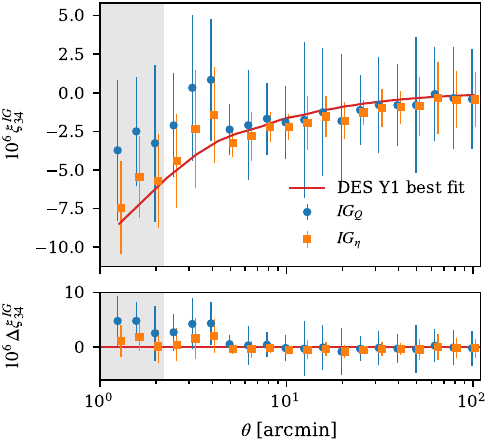}
    \caption{The intrinsic alignment - cosmic shear signals between bins 3 and 4. Illustrated is the theoretical prediction obtained from CCL using the IA parameters obtained from DES year 1 (solid red). We also plotted 2 different uses of the scaling relation (Equation \ref{eq:scalingrelation}): using  $\eta$ for the separating out the initial $Ig$ signal (solid orange) and using $Q_i$ (solid blue). We have also plotted the difference between the detected signal and the CCL estimated signal (lower).}
    \label{fig:IG}
\end{figure}

Finally, we obtained the intrinsic alignment-cosmic shear cross correlation between bins. We did this by using the scaling relation given in Equation (\ref{eq:scalingrelation}) and the galaxy bias estimate obtained from the $w^{Gg}$ correlations.
Of all the terms in Equation (\ref{eq:cosmicshear}), we expect the $\braket{\gamma^I_i, \gamma^G_j}$ to be the major intrinsic alignment contributor when we are looking at cross-bin correlations. From the results in the previous section, we can find the potential cross-bin two-point correlation for intrinsic alignment. For the various bin combinations with $i<j$, while we do compute all relevant bin combinations we will here focus on the bin combination with the highest Signal-to-Noise ratio (bin combination 3,4), see Appendix \ref{app:SNR} for more details. To estimate the galaxy bias $\tilde{b}$ in Equation (\ref{eq:scalingrelation}) we utilize the $Gg$ signal found earlier as described in Equation (\ref{eq:bias_Gg}). We show the result of our chosen bin combination in Figure (\ref{fig:IG}). It is important to note that the cosmic shear term we obtain here, isn't the intrinsic alignment term of either $\xi_\pm$, but rather a linear combination of the intrinsic alignment terms. 

We compare this Cosmic shear intrinsic alignment signal to a null detection, and we find that for the $Q$ version we get $5.1\times10^{-7}\sigma$ and the $\eta$ version we get $0.237\sigma$, while if we compare with the theoretical prediction we get for $Q$ a reduced $\chi^2 = 0.125$ and for $\eta$ we get reduced $\chi^2 = 0.0626$, which suggest that while our data looks similar to the model, the uncertainties are too large to make any additional statements.

\section{Discussion \& Conclusion}
\label{sec:Discussion}

\subsection{Photometry issues}
We have found several issues that we believe are caused by what essentially boils down to a classic physics issue: "that which fits an ensemble does not always fit the individual elements that make up the ensemble". What is meant by this is is as follows: \citet{Hoyle_2018} found that what really mattered to the cosmological weak lensing analysis that DES was designed for, was the ensemble averages and other metrics of the ensemble, but not the shape of the individual galaxy redshift probability. The same conclusion cannot be drawn for our self-calibration method: we specifically care deeply about the shape of our redshift distribution, both in photometric redshift space and in true redshift space. But, more importantly, we need to ensure that the correspondence between these two PDFs reflects the actual galaxies and not just their ensembles. As we noted in Section \ref{sec:PRD} we are limited in how we can recover the PDF. However, we are dominated by the fact that this PDF is for the ensembles and not a combination of individual galaxies distributions. As can been seen in Figure \ref{fig:redshift_dists}, the same set of galaxies can have different redshift distributions depending on how we try to recreate the PDF. This also means that when we try to estimate $(\eta)$ that should tell us how well we can within the bin distinguish between galaxies along the line of sight, the different recreations doesn't always agree; this is well illustrated in Figure \ref{fig:eta_bins} where we see that for bin 2, that the $\eta$s estimated from the different PDF recreations are quite different inside the bin range. In Table \ref{tab:eta_estimators} and Figure \ref{fig:Q_of}, we see that the different recreations of the PDF changes the estimates we get for the Quality parameter ($Q$) or it's approximation $\eta$. This uncertainty in the Quality parameter is in our view a point of challenge of our method, and represents a significant limitation for the method. 

This also leads us to question our actual ability to rely on the point estimates for the individual galaxies to estimate the two correlations given by Equation (\ref{eq:wgammag}) and Equation (\ref{eq:wgammagS}). Most weak lensing surveys like DES find that the individual galaxies point-estimates are not as important as the constraints on the distribution \citep{Hoyle_2018}. This lack of reliable point-estimates could help explain why in Figure \ref{fig:gamT} the two correlations seem not well separated for all but the third bin. The third bin also happens to be the bin that \citet{Hoyle_2018}'s analysis found to be the most reliable of the bins.

This means that while we find some indication of an IA signal in the third and fourth bins of DES year 1, there is still a long way to go for the signal to be robust and that we can for sure say it is not just noise. We found that the recreation of the PDFs and their use in calculating $\eta$ can easily mislead one to believe they have much better separation along the line of sight than what is actually achievable with the data. We find that for at least the best bin (the third) the $\eta$ seems to converge towards a similar value, but this does not translate well into the $Q$'s estimated from them (see Figure \ref{fig:Q_of} and Table \ref{tab:eta_estimators}). So, in some sense, the $\eta$ could be viewed as a rougher but more reliable estimate than the $Q$s since the $Q$s seem even more susceptible to variations in the redshift distributions. 

It is our hope that in future surveys, the need for individual galaxies photo-$z$ information is recognized and that that will make investigations such as ours more viable. We can, however, see that at least at the ensemble level, $\eta$ can be stable when we are focusing on the bin that seems to have the least photo-$z$ issues. This result is similar to the one we found in \citet{Pedersen:2019wfp} for KiDS-450.  
The dependencies of the detection on the quality of the photometry is being investigated by \citet{Buzzard_SC} , where $Q$ is found to be a reliable predictor for the success of the self calibration algorithm to separate the $gI$ and $gG$ signal accurately. 

Regarding why the 3rd bin would work well, our hypothesis is the redshift bin least plagued by uncertainties in photo-zs. It is has been seen in the past that at lower redshifts photo-$z$ estimators struggle with outliers especially around 0.45 in redshift caused by the Balmer break being in completely inside the $g$ band \citep{Kalmbach_2020, Graham_2020}. However at the redshift range corresponding to the 3rd bin the Balmer break is now starting to move into the $r$ band and therefore a much more accurate determination of it is possible and that leads to better photo-$z$s. 

\subsection{Separating Ig and Gg}
When it comes to separating the signals we can look at Table (\ref{tab:MD_Gg_Ig}) and note that only the 3rd bin has a separation of the $Ig$ and $Gg$ signal that is above $5\sigma$, which would suggest we are struggling to reliably separate the two signals for most of our bins. 
As noted in Section \ref{sec:Ig} we only find an intrinsic alignment signal that is non-zero in the last two of the redshift bins. While we find that the intrinsic alignment signal in those bins is also in some contention with the theoretically predicted signal. We believe a large part of this is due to our photo-z issues as outlined above, but it is interesting to note that the later DES year 3 analysis also couldn't rule out a null intrinsic alignment signal \citep{DESy3}.

As we would expect the isolated $Gg$ signal as shown in Figure \ref{fig:IgGg} aligns fairly well with the theoretical expectations, see Table \ref{tab:Ggchi2}. While there seems to be more issues between the $Ig$ signal and the NLA model, see Table \ref{tab:Igchi2}.

\begin{table}
    \centering
    \caption{the reduced $\chi^2$ values for comparing the $w^{Gg}$ with the theoretical prediction of the same. When comparing these numbers with Figure \ref{fig:IgGg} the $\chi^2$ here in the first column is comparing the solid lines, and the second column is comparing the dashed lines.}
    \label{tab:Ggchi2}
    \begin{tabular}{ccc}
    \hline
bin & $\chi^2 (Q)$ & $\chi^2 (\eta)$  \\
\hline
1 & 0.258 & 0.258 \\ 
2 & 0.451 & 0.433 \\ 
3 & 0.788 & 0.768 \\ 
4 & 0.426 & 0.525 \\ 
\hline
    \end{tabular}
\end{table}

\begin{table}
    \centering
    \caption{the reduced $\chi^2$ values for comparing the $w^{Ig}$ with the theoretical NLA prediction of the same. When comparing these numbers with Figure \ref{fig:IgGg} the $\chi^2$ here in the first column is comparing the solid lines, and the second column is comparing the dashed lines.}
    \label{tab:Igchi2}
    \begin{tabular}{ccc}
    \hline
     bin & $\chi^2 (Q)$ & $\chi^2 (\eta)$  \\
     \hline
1 & 0.92 & 0.263 \\ 
2 & 1.16 & 0.372 \\ 
3 & 1.13 & 1.04 \\ 
4 & 0.727 & 0.713 \\
    \hline
    \end{tabular}
    
\end{table}

\subsection{Obtaining the IG correlation}
As can be seen in Figure \ref{fig:IG}, we can translate the $Ig$ of bin 3 into an $IG$ signal for the correlation of bin 3 and bin 4. This signal does look similar to the theoretical prediction, but as noted before, we have issues with the lack of constraints on galaxy bias. In a better scenario, better constrains of the galaxy bias in the source bins could be developed and that will make this more reliable. We have tried also propagate the covariance between the $Ig$ and the $b_i$ signal into the $IG$, but since both signals are obtained from the two initial correlations, it would have been good to find further evidence that galaxy bias values used are in fact reliable.

\subsection{Other concerns}

We also note that there is the potential in redshift weighting of the selected correlations versus the non-selected correlations, in the form of an uneven redshift distribution within each bin. This was investigated in \cite{Yao:2020jpj} where the authors found this effect to seemingly cancel itself out. All of this relies on the scaling to still fit in the thin bin approximation, which we believe it still does. But we leave for future simulation work to evaluate the comprehensive effects of bin size on the ability to isolate the IA signal. 

Another worry that arose while discussing this paper, was that the fact that we are trying to use source galaxies as density tracers, and this is can be tricky for purely photometric source samples, due to inhomogeneities in the surveys \citep{Leistedt_2013,Leistedt_2014,Yan_2025}. This could also be causing issues, and we leave it for future work to explore the effects these inhomogeneities in the survey could have on the self-calibration method.

Looking at both the $Ig$ and $IG$ correlations, we note that at larger separation there is good agreement between the NLA-alignment model and the signal we found. However, at smaller separation there is some suggestion that the best fit overestimates the amplitude of the contamination. This is consistent with the state of the field where modeling IA in the non-linear regime remains an active and challenging area of work \citep{Blazek:2017wbz, Vlah:2020ovg}

\subsection{Future prospects}
\label{sec:future}
Looking forward, our intent is to construct a similar reanalysis of HSC data and the KiDS 1000 set. However, before we do this, work is required on improving the integration of photo-z codes with TXPipe and, especially, with our extension to it, such that we can obtain more reliable estimates of the PDF taking into account the issues outlined above. 

A further requirement, which we are developing for the HSC and KiDS reanalyses mentioned above, is the treatment of cosmic magnification (Appendix~\ref{app:magnification}). Because it contaminates the self-calibrated IA signal and grows in importance at the redshifts probed by LSST, we regard it as a necessary component of any Stage-IV self-calibration analysis. We intend to incorporate into the pipeline a catalogue-level estimate of the magnification coefficient that propagates the full source selection, a magnification template in the IA model, and a magnitude-weighting cross-check, together with random catalogues for the source density field, and to present this in a future publication.
The other limiting factor in this paper is the lack of constraints on the galaxy bias for source galaxies. This factor is caused by lack of a proper random catalogs for source galaxies for the DES source distribution. This comes from the complexity of generating these catalogs, basically requiring survey-style simulations. Another issue regarding random catalogs is simply a question of size. We expect random catalogs to be anywhere from 8 to 100 times larger than the original catalog. This means that for a catalog such as the DES year 1 with $\approx26$ million source galaxies, we would need to have almost 100 million objects in a random catalog, which would put significant stress on current computing infrastructure. Alternative methods might be possible, see for example the discussion of random catalogs in \cite{TXPipe2022}, maybe something similar is possible for sources. Other interesting avenues to approach regarding the galaxy bias is to further investigate the implications of second order bias, beyond what we did in Appendix \ref{app:IABias}.

In this paper, we have presented the first implementation of the LSST-DESC's self-calibration of intrinsic alignment software. This will serve as an extension on the TXPipe 3x2 point software being designed for analysis of the upcoming LSST. This is also a first application of such a method to the DES Y1 data. We intend to apply this pipelines software on other precursor surveys and simulated data sets as verification of it's methodology, and we also expect that we can use this for better constrains of intrinsic alignment models. This work fits within the overall goal of preparing the pipeline for the incoming Rubin LSST data and its analysis within the DESC collaboration.

\section*{acknowledgments}
This paper has undergone internal review in the LSST Dark Energy Science Collaboration.
EP, LM, and all the rest of the authors would like to thank Elisa Chisari and Jonathan Blazek for their thorough for many helpful and productive discussions. All the authors would also like to thank Dr. Ji Yao for his thoughtful feedback as reviewer. 
EP would like to thank Chris Stubbs at Harvard and acknowledge the support given by the DOE via the Cosmic Frontier grant DE-SC0007881. 
MI acknowledges that this material is based upon work supported in part by the Department of Energy, Office of Science, under Award Number DE-SC0022184 and also in part by the U.S. National Science Foundation under grant AST2327245.
EP and DL would also like to acknowledge the support from by the Science and Technology Facilities Council (STFC) [grant No. UKRI1172].
This project used public archival data from the Dark Energy Survey (DES). Funding for the DES Projects has been provided by the U.S. Department of Energy, the U.S. National Science Foundation, the Ministry of Science and Education of Spain, the Science and Technology FacilitiesCouncil of the United Kingdom, the Higher Education Funding Council for England, the National Center for Supercomputing Applications at the University of Illinois at Urbana-Champaign, the Kavli Institute of Cosmological Physics at the University of Chicago, the Center for Cosmology and Astro-Particle Physics at the Ohio State University, the Mitchell Institute for Fundamental Physics and Astronomy at Texas A\&M University, Financiadora de Estudos e Projetos, Funda{\c c}{\~a}o Carlos Chagas Filho de Amparo {\`a} Pesquisa do Estado do Rio de Janeiro, Conselho Nacional de Desenvolvimento Cient{\'i}fico e Tecnol{\'o}gico and the Minist{\'e}rio da Ci{\^e}ncia, Tecnologia e Inova{\c c}{\~a}o, the Deutsche Forschungsgemeinschaft, and the Collaborating Institutions in the Dark Energy Survey.
The Collaborating Institutions are Argonne National Laboratory, the University of California at Santa Cruz, the University of Cambridge, Centro de Investigaciones Energ{\'e}ticas, Medioambientales y Tecnol{\'o}gicas-Madrid, the University of Chicago, University College London, the DES-Brazil Consortium, the University of Edinburgh, the Eidgen{\"o}ssische Technische Hochschule (ETH) Z{\"u}rich,  Fermi National Accelerator Laboratory, the University of Illinois at Urbana-Champaign, the Institut de Ci{\`e}ncies de l'Espai (IEEC/CSIC), the Institut de F{\'i}sica d'Altes Energies, Lawrence Berkeley National Laboratory, the Ludwig-Maximilians Universit{\"a}t M{\"u}nchen and the associated Excellence Cluster Universe, the University of Michigan, the National Optical Astronomy Observatory, the University of Nottingham, The Ohio State University, the OzDES Membership Consortium, the University of Pennsylvania, the University of Portsmouth, SLAC National Accelerator Laboratory, Stanford University, the University of Sussex, and Texas A\&M University.
Based in part on observations at Cerro Tololo Inter-American Observatory, National Optical Astronomy Observatory, which is operated by the Association of Universities for Research in Astronomy (AURA) under a cooperative agreement with the National Science Foundation.

Contributions:
EP: Wrote the main parts of the paper. Conducted the analysis. Wrote the extension for TXPipe. Wrote the other code used for the analysis.
LM: Wrote second pipeline, helped rewriting and editing the paper.
EPL: Co-wrote code, helped with analysis and data-reduction.
MI: Supervising PI, contributed to various phases of the analysis, co-wrote the paper.
JZ: DESC Builder, Co-wrote primary data code.
CC: Co-wrote code, helped with analysis, and paper.
DL: Review and improving the narrative of the paper, editing.

\section*{Data Availability}
As mentioned above we used publicly available public archival data from the Dark Energy Survey (DES), which can be found at \url{https://des.ncsa.illinois.edu/releases/y1a1}. The interim data products will be shared upon reasonable request to the corresponding author. 
The TXPipe code base can be found at \url{https://github.com/LSSTDESC/TXPipe}, while the SCIA2020 will be shared upon reasonable request to the corresponding author. We hope to release the latter to the public soon.
%%%%%%%%%%%%%%%%%%%% REFERENCES %%%%%%%%%%%%%%%%%%

% The best way to enter references is to use BibTeX:

\bibliographystyle{mnras}
\bibliography{bibliography} % if your bibtex file is called example.bib

% Alternatively you could enter them by hand, like this:
% This method is tedious and prone to error if you have lots of references
%\begin{thebibliography}{99}
%\bibitem[\protect\citeauthoryear{Author}{2012}]{Author2012}
%Author A.~N., 2013, Journal of Improbable Astronomy, 1, 1
%\bibitem[\protect\citeauthoryear{Others}{2013}]{Others2013}
%Others S., 2012, Journal of Interesting Stuff, 17, 198
%\end{thebibliography}

%%%%%%%%%%%%%%%%%%%%%%%%%%%%%%%%%%%%%%%%%%%%%%%%%%

%%%%%%%%%%%%%%%%% APPENDICES %%%%%%%%%%%%%%%%%%%%%

\appendix

\section{Intrinsic alignment and galaxy bias}
\label{app:IABias}
In this section we will expand on the second term of Equation (\ref{eq:galaxygalaxylensing}). There are several ideas; the main idea is that we theoretically can predict the underlying matter-powerspectra, represented here by the matter over-density $\left(\delta_m\right)$. However, how is this related to the two components of $\braket{\gamma^I_i, \delta_i^g}$? There is a large number of theories on this, and especially how $\gamma^I_i$\footnote{We focus here only on the $E$ component of the field.} looks is a very active field. Here we will follow the notation of \cite{Blazek:2017wbz} and we will describe the intrinsic alignment field as a series of terms in orders of the matter density:
\begin{align}
    \gamma^I &= C_1 f_E \delta_m(k) + c_{1\delta} \int \frac{d^3 k_1}{\left(2\pi\right)^3}f_E (k_1) \delta_m (k_1) \delta_m (k_2) \\
    &\quad + C_2\int \frac{d^3k_1}{\left(2\pi\right)^3}\frac{d^3k_2}{\left(2\pi\right)^3}h_E(k_1,k_2) \delta_m(k_1)\delta_m(k_2) \delta_m(k_3) + \ldots 
\end{align}
So we see that to first order the intrinsic alignment field is expected to be linear with the matter density. Let us also remind the reader that the reason we are interested in the shear field initially is that we expect $\gamma^G \propto \delta_m$.

Next let us discuss how we can expand $\delta_i^g$ in terms of $\delta_m$, it turns out that we expect the galaxies to have a sort of bias in its proportionality to the matter powerspectra. We describe it here analogous with \cite{Pandey:2020zyr}:
\begin{equation}
    \delta_g = b_1 \delta_m + \frac{b_2}{2}\left(\delta^2 - \braket{\delta_m}^2\right) + \frac{b_s}{2}\left( s^2 - \braket{s}^2\right) + \ldots 
\end{equation}
Here we have to introduce a second field in the form of the $s$ which represents the tidal field, defined in \cite{2009JCAP...08..020M}.

Now returning to $\braket{\gamma^I_i, \delta_i^g}$ we can see that there are several different combinations possible. It turns out as long as the bias is assumed linear we don't have to worry about the order of the intrinsic alignment field. This was what was done in \cite{Pedersen:2019wfp}. If we go to non-linear bias the picture is not as clear, we can start by looking at the linear term for the intrinsic alignment, in this case, we get the for the shear - intrinsic alignment term:
\begin{equation}
    \braket{\gamma^G \gamma^I} \propto  c_1 f_E \braket{\delta_m \delta_m}
\end{equation}
while for the galaxy - intrinsic alignment term:
\begin{align}
    \braket{\delta_g \gamma^I} &= c_1 f_E \left[ b_1 \braket{\delta_m \delta_m} + \frac{b_2}{2}\left(\braket{\delta_m^2 \delta_m} - \braket{\braket{\delta_m}^2 \delta_m}\right) \right.\\
    &\quad\left.+ \frac{b_s}{2}\left(\braket{s^2 \delta_m} - \braket{\braket{s}^2 \delta_m}\right)\right]
\end{align}
This is what we will be assuming to use for our study here, in this case we see that the ratio needed for obtaining the scaling relation (\ref{eq:scalingrelation}), becomes:

\begin{equation}
    \frac{\braket{\delta_g \gamma^I}}{\braket{\delta_m \gamma^I}} = \frac{b_1 \braket{\delta_m \delta_m} + \frac{b_2}{2}\left(\braket{\delta_m^2 \delta_m} - \braket{\braket{\delta_m}^2 \delta_m}\right)+ \ldots}{\braket{\delta_m \delta_m}}
\end{equation}
which incidentally is the same as the ratio $\frac{\xi_{gm}}{\xi_{mm}}$ used in \cite{Pandey:2020zyr}.
The work presented in this appendix was in parallel to this paper expanded upon in \citep{Bera:2025ixc}.

\section{Independent pipeline}
\label{app:Secondpipe}
An independent pipeline was developed to reproduce the results obtained from TXPipe and SCIA2020. This pipeline is spread out in 5 pieces of code that complete the following tasks: 

\begin{itemize}
\item{Estimating the photo-$z$ PDF from the Monte Carlo realizations in the DES data}
\item{Integrating the $\eta$ parameter using the photo-$z$ PDF}
\item{Integrating the $Q$ parameter using the photo-$z$ PDF}
\item{Binning and applying the Metacalibration algorithm to the shear catalog}
\item{Using the Metacalibrated shears, in conjunction with information from the photo-$z$ PDF to calculate the relevant correlation functions}
\end{itemize}

Results agree with TXPipe very well. \\
The metacalibration algorithm is described in \cite{Zuntz_2018}. This is applied after the binning, separately for each redshift bin, using the quality flags provided in the DES data in conjunction with the redMaGiC mask, such that each bin has its own $S$ and $R$ selection and response matrices. $R$ was double checked with the values provided within the DES data to be the same for the same bin cuts.

\section{Cosmology dependence in the $Q$ parameter}
\label{app:Qcosmo}
\begin{figure}
    \centering
    \includegraphics[width=\columnwidth]{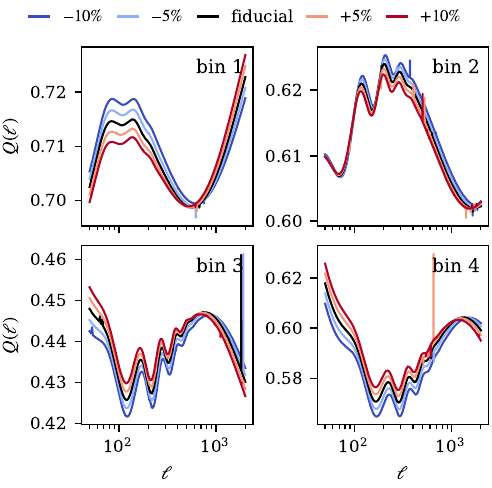}
    \caption{Comparison of difference cosmological values and their effects on $Q$. In blue is the "regular" cosmology, i.e. unchanged from the DES y1 best fit cosmology. while the green and orange lines are the same cosmology with $\pm5\%$ changes for $h_0$ and $\sigma_8$. The red and blue lines are then with $\pm10\%$ changes to $h_0$ and $\sigma_8$.}
    \label{fig:Q_cosmo}
\end{figure}

We ran a test, where we varied the cosmology parameters when we were estimating the $Q$ from the $\eta$ using CCL. We used the DES best fit cosmology, referred to as "regular", then we did the same but changed $h_0$ and $\sigma_8$ by $\pm5\%$ (high and low) and by $\pm10\%$ (high 2 and low 2). The results of this are shown in Figure \ref{fig:Q_cosmo}. For the underlying $\eta$ and redshift distribution we used the Gaussian method outlined in Section \ref{sec:Results}.

As we can see we see some variation in $Q$ dependent on the two cosmological parameters, however we note that the changes are quite small, for both $Q\left(\ell\right)$ and $Q\left(\theta\right)$ we find that the changes compared to the baseline is rarely more $\pm 0.02$(in absolute numbers) and the higher deviations in $h_0$ and $\sigma_8$ resulting in the higher deviations. but a $0.02$ swing in $Q$, compared to the actual values is about a factor 10 smaller than the variations we see in just comparing the $\eta$s between methods in Table \ref{tab:eta_estimators}.

\section{Signal to noise discussion}
\label{app:SNR}
Another possible method of investigating the stability of the results would be to do a Signal-to-Noise ratio (SNR) for our measurements. We have calculated the signal to noise ratio for each of the calculated steps, comparing the data with the size of the estimated errors. To estimate the SNR we calculate a few different estimators for it:
\begin{itemize}
    \item Mean SNR, by which we mean: \begin{equation}
        SNR_\text{mean} = \frac{1}{N}\sum_i^N \frac{x_i}{\sigma_i}.
    \end{equation}
    where $x_i$ is the data-value for each individual data point.
    \item Median SNR, i.e. as above we compute the signal-to-noise for each data point, and take the median.
    \item the RMS (root-mean-square) SNR:
    \begin{equation}
        SNR_\text{RMS} = \frac{\sqrt{\frac{1}{N}\sum \left(x_i -\bar{x}\right)^2}}{\bar{\sigma}}
    \end{equation}
\end{itemize}
We summarize the SNR for the $Ig$ signals from the four bins in Table \ref{tab:SNR_Ig}. It can be seen that almost none of the signal to noise ratios rise above 1, though an upward trend is observed in the higher redshift bins.
\begin{table}
    \centering
    \begin{tabular}{cccc}
    \hline
    bin & $SNR_\text{mean}$ & $SNR_\text{med}$ & $SNR_\text{RMS}$ \\
    \hline
      1 (using $Q$)   &  0.611 & 0.275 & 0.673\\
      1 (using $\eta$) & 0.282 & 0.128 & 0.421 \\
      2 (using $Q$) & 0.745 & 0.707 & 0.861 \\
      2 (using $\eta$) & 0.374 & 0.374 & 0.909 \\
      3 (using $Q$) & 0.937 & 0.868 & 0.532 \\
      3 (using $\eta$) & 0.711 & 0.658 & 0.803 \\
      4 (using $Q$) & 0.889 & 0.719 & 1.06 \\
      4 (using $\eta$) & 0.83 & 0.673 & 1.07\\
      \hline
    \end{tabular}
    \caption{Signal-to-noise ratios for the separated $Ig$ signals for all four bins, using both $\eta$ and $Q$.}
    \label{tab:SNR_Ig}
\end{table}

\begin{figure*}
    \centering
    \includegraphics[width=\textwidth]{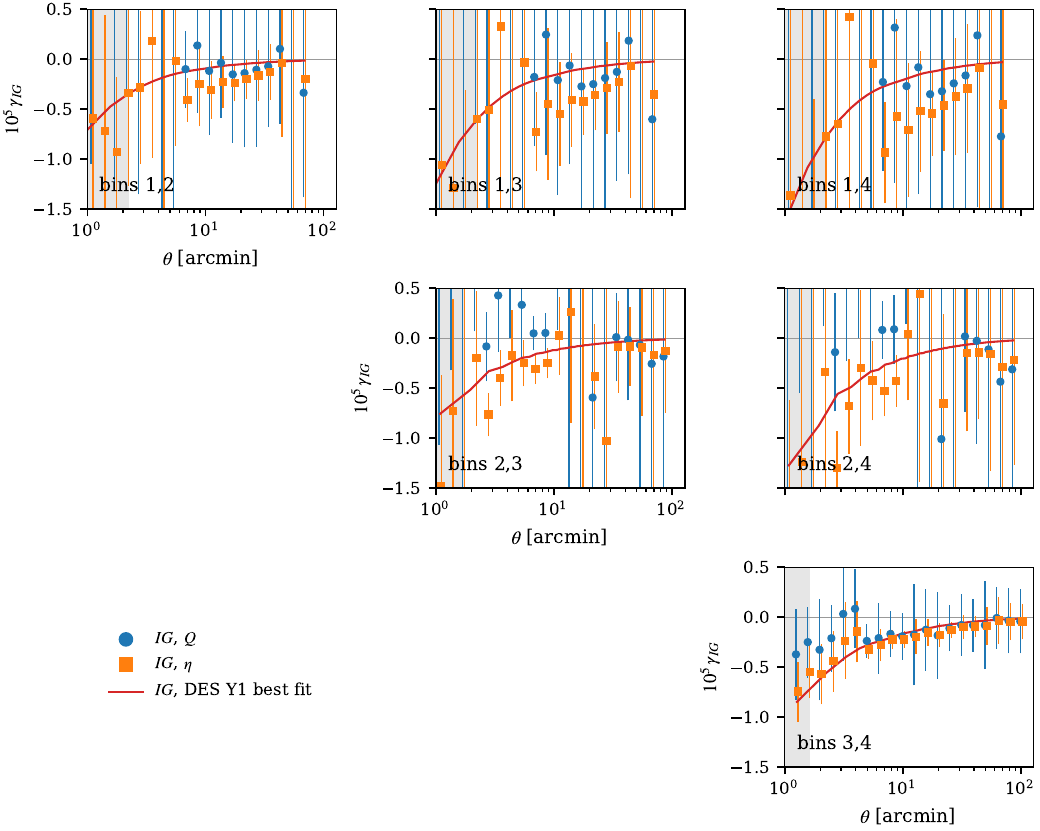}
    \caption{The $IG$ signal, coming from the scaling relation (Equation \ref{eq:scalingrelation}). The green line is the theoretical signal estimated using the NLA method as outlined in Section \ref{sec:IA}, the blue datapoints is the scaled signal using $Q$, and orange datapoints is using $\eta$. The grey area is the cut data as discussed previously, the cut data is excluded from the estimations of $SNR$, listed in Table \ref{tab:SNR_Ig}.}
    \label{fig:IG_SNR}
\end{figure*}

We also can look at how the scaled $IG$ signal behaves in Figure \ref{fig:IG_SNR}. It is clear that only the last bin combination (3,4) have a SNR even close 1. This points to our results being noise/uncertainty dominated.  

If we focus on the $\eta$ separated signal, we notice that only the last two bins exceed a 0.5 $SNR$, which suggests we are noise dominated. To summarize, the SNR from the $gI$ signal separation utilizing $Q$  is close to 1 in the higher redshift bins, while the extra uncertainties propagation applied through the scaling relation for $GI$  make it perform worse. 

\section{PDF realizations}
\begin{figure*}
    \centering
    \includegraphics[width=\textwidth]{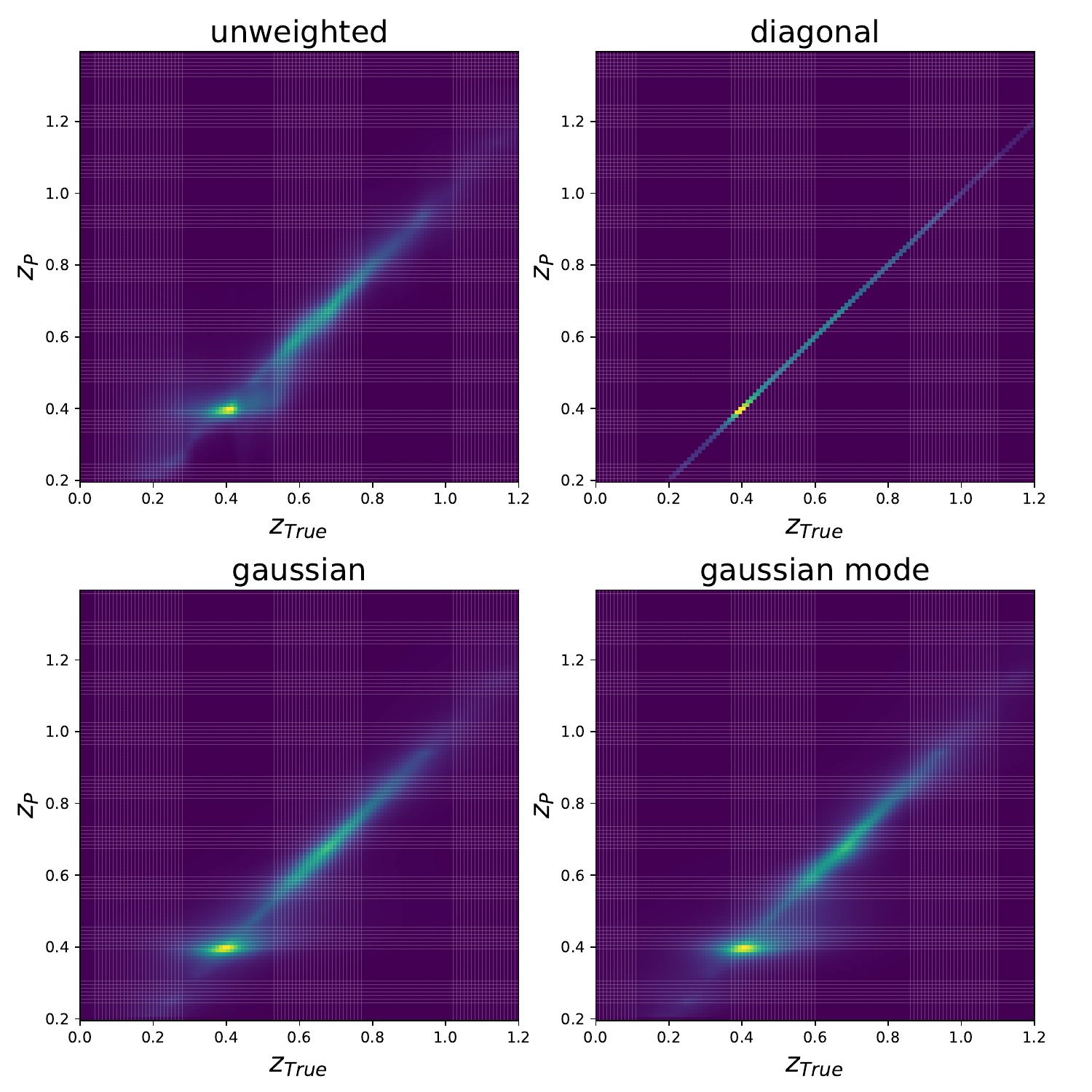}
    \caption{The 4 different realizations of the combined probability distributions for the DES Y1 dataset. On the x-axis is the assumed true redshift, while the y-axis is the reported photo-z.}
    \label{fig:PDFS}
\end{figure*}

For illustrative purposes we show here the different possible realization of PDFs we have used, they are all shown in Figure \ref{fig:PDFS}.

\section{Double $Q$ formalism}
\label{app:doubleQ}
It was suggested in \citet{Yao:2020jpj} that a way to overcome some of the issues of the uneven distribution within bins would be to introduce a second $Q$, such that:
\begin{align}
    Q_{Gg} &= \frac{\left.w^{Gg}\right|_S}{w^{Gg}} \\
    Q_{Ig} &= \frac{\left.w^{Ig}\right|_S}{w^{Ig}}
\end{align}
So we see the first is the same as the $Q$ given by Equation \ref{eq:Q}. The second $Q$ $(Q_{Ig})$ is also estimated via an $\eta$, however in this case we omit the Lensing kernel from Equation \ref{eq:eta}. 

This also means a change to the separation Equations \ref{eq:wGg2},\ref{eq:wIg2} to instead be:
\begin{align}
    w^{Gg}_{ii} &= \frac{Q_{Ig,i}\times w^{\gamma g}_{ii}\left(r_p\right) - \left.w^{\gamma g}_{ii}\right|_S\left(r_p\right)}{Q_{Ig,i}-Q_{Gg,i}} \\
    w^{Ig}_{ii} &= \frac{\left.w^{\gamma g}_{ii}\right|_S\left(r_p\right) - Q_{Gg,i}\times w^{\gamma g}_{ii}\left(r_p\right)}{Q_{Ig,i}-Q_{Gg,i}}.
\end{align}

\begin{figure*}
    \centering
    \includegraphics[width=\textwidth]{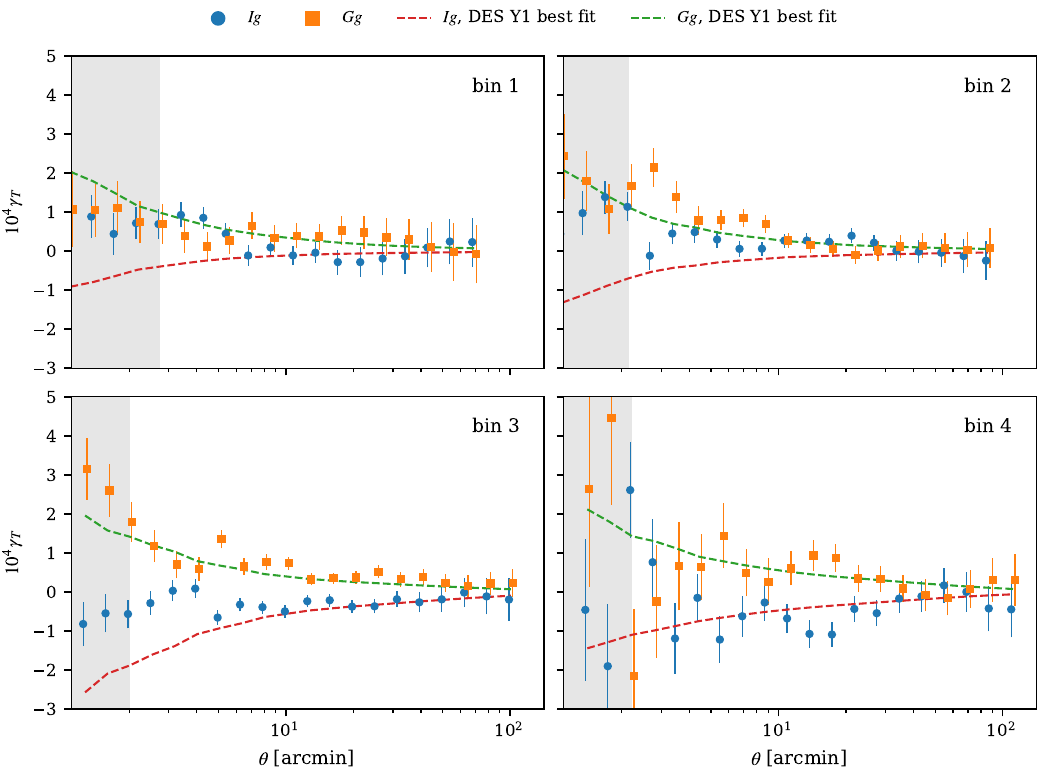}
    \caption{Ig Gg separation, for all 4 bins, similar to Figure \ref{fig:IgGg}. We have not done a $\eta$ version with the double $\eta$s. The theoretical lines represent the theoretical estimates for each bin, calculated using CCL, and the galaxy bias as described previously.}
    \label{fig:double_Q_sep}
\end{figure*}

We calculated the two new $Q$'s for the dataset; we list the mean $Q$ in table \ref{tab:QQ}. As can be seen, we don't find much effect in the $Q_{Ig}$. We find about the same levels of detection as the single $Q$ method, see Figure \ref{fig:double_Q_sep}. In fact only real change we saw was that the comparison between $Ig$ and the best fit, went from a $3\sigma$ to $3.6\sigma$. This is of course in agreement with \citet{DESy3} results, that also suggested a null result. 

\begin{table}
    \centering
    \begin{tabular}{ccc}
    \hline
    bin & $Q_{Gg}$ & $Q_{Ig}$ \\
    \hline
       1  & 0.703 & 0.978\\
       2  & 0.610 & 1.068\\
       3  & 0.448 & 1.004\\
       4  & 0.618 & 0.994\\
    \hline 
    \end{tabular}
    \caption{Mean $Q$ values for the double $Q$ formalism.}
    \label{tab:QQ}
\end{table}

\subsection{Cosmic magnification}
\label{app:magnification}
The double-$Q$ formalism also makes explicit how cosmic magnification enters the separation. Weak lensing by foreground structure modulates the observed number density of the galaxies used as density tracers \citep{BartelmannSchneider_2001},
\begin{equation}
    \delta_g^\text{obs} = \delta_g + g_\text{mag}\,\kappa_g,
    \qquad g_\text{mag} = 2(\alpha-1),
\end{equation}
where $\kappa_g$ is the convergence of the density sample and $\alpha$ the effective slope of its cumulative number counts. Both observables then acquire a shear--convergence term $w^{G\kappa}_{ii}$, with its own selection ratio $Q_{G\kappa,i}$:
\begin{align}
    w^{\gamma g}_{ii}\left(r_p\right) &= w^{Gg}_{ii}\left(r_p\right) + w^{Ig}_{ii}\left(r_p\right) + g_\text{mag}\,w^{G\kappa}_{ii}\left(r_p\right), \\
    \left.w^{\gamma g}_{ii}\right|_S\left(r_p\right) &= Q_{Gg,i}\,w^{Gg}_{ii}\left(r_p\right) + Q_{Ig,i}\,w^{Ig}_{ii}\left(r_p\right) \nonumber \\
    &\quad + g_\text{mag}\,Q_{G\kappa,i}\,w^{G\kappa}_{ii}\left(r_p\right).
\end{align}
Inserting these into the separation equations above, the separated signals are
\begin{align}
    w^{Gg}_{ii} &\rightarrow w^{Gg}_{ii} + g_\text{mag}\,\frac{Q_{Ig,i}-Q_{G\kappa,i}}{Q_{Ig,i}-Q_{Gg,i}}\,w^{G\kappa}_{ii}, \\
    w^{Ig}_{ii} &\rightarrow w^{Ig}_{ii} + g_\text{mag}\,\frac{Q_{G\kappa,i}-Q_{Gg,i}}{Q_{Ig,i}-Q_{Gg,i}}\,w^{G\kappa}_{ii}.
\end{align}
Because $\kappa_g$, like the IA term, does not depend on the relative ordering of the pair along the line of sight, $Q_{G\kappa,i}\simeq Q_{Ig,i}$ \citep{Yao2023}: the magnification term drops out of $w^{Gg}_{ii}$ and enters $w^{Ig}_{ii}$ with its full amplitude. The single-$Q$ separation of Equations (\ref{eq:wGg2}) and (\ref{eq:wIg2}) is the special case $Q_{Ig,i}=Q_{G\kappa,i}=1$ and leads to the same conclusion. The near-unity values of $Q_{Ig}$ in Table \ref{tab:QQ} therefore imply that, for our sample, the separated $Ig$ signal carries essentially the full magnification contribution.

The relative importance of this contamination grows with redshift, because $w^{G\kappa}_{ii}$ accumulates lensing along the whole line of sight whereas $w^{Ig}_{ii}$ is local. \citet{Yao2023} find that neglecting it biases the IA amplitude by $\sim0.3$--$0.4$ at $\bar z\simeq0.8$--$0.9$, a large fraction of the IA amplitude inferred by DES Y1 \citep{DES:2017qwj}. Our third and fourth bins lie in this regime or beyond, and the redshift distribution expected for LSST makes the problem more severe. The sign of the effect depends on the sign of $g_\text{mag}$: since $w^{Ig}_{ii}$ is negative for radial alignments and $w^{G\kappa}_{ii}$ is positive, a negative $g_\text{mag}$ mimics intrinsic alignment and a positive one partially cancels it. The indications of an IA signal we report in bins 3 and 4 therefore cannot be attributed to intrinsic alignment alone until this term is quantified.

Quantifying it requires the magnification coefficient of the source sample, which we have not measured here. For a sample defined by photometric-redshift binning together with signal-to-noise, size and shape-measurement cuts, as for the DES Y1 Metacalibration sample, magnification moves galaxies across all of these selection boundaries, and the effective coefficient can differ substantially from the value implied by the flux limit alone \citep{Qin2025}. It must therefore be estimated by propagating the complete selection. Because the coefficient depends on the shape of the luminosity function near the selection limits, it is also expected to differ between galaxy populations. This is relevant for colour-split tests of the separated signal, a natural validation of self-calibration, since a magnification contribution differing between subsamples could mimic or mask a genuine difference in alignment.

Two complementary mitigation strategies exist. The magnification term can be modelled, adding $g_\text{mag}w^{G\kappa}_{ii}$ to the IA model with $g_\text{mag}$ as a nuisance parameter \citep{Yao2023}; the resulting degeneracy with the IA amplitude can be reduced by an independent, catalogue-level estimate of $g_\text{mag}$. Alternatively, since positions and shapes enter the estimator as separate catalogues, the density tracer alone can be reweighted in magnitude bins so that its effective magnification response vanishes, removing the term by construction \citep{Qin2024,Qin2025}, at the cost of increased shot noise and a modified effective redshift distribution of the density sample, which would then require $Q$ to be recomputed. We leave a full treatment of magnification and these related effects to future work.

\section{Theoretical signal size}
\label{app:signal_size}
\begin{figure}
    \centering
    \includegraphics[width=\columnwidth]{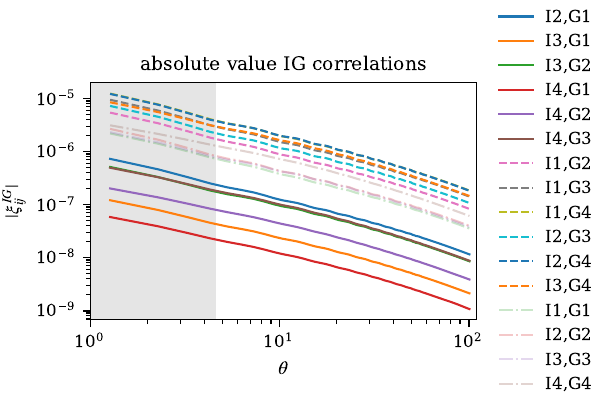}
    \caption{Theoretical estimates of the $IG/ GI$ signal, using the actual redshift distributions and the NLA model. There are 3 groups of signals, those for the $GI$, where the $G$ bin is in front of the $I$ bin (solid lines), the opposite case $IG$ (dashed lines), and finally the "auto-corrolations" (dashed faint lines). the y-axis is logarithmic values of the absolute value of the signal, so we can see that the solid lines are all smaller by aleast a magnitude.}
    \label{fig:signalstrength}
\end{figure}

To estimate how well our expectation that we can ignore the $\braket{\gamma^G_i, \gamma^I_j}$ term, we have using the redshift distributions in Figure \ref{fig:redshift_dists} estimated the theoretical signal size for each combination, we illustrate this in Figure 
\ref{fig:signalstrength}, where we can see all the signals we had initially discounted are indeed much smaller signals compared to the ones we were hoping to detect. This does raise the question about if we look at very detailed surveys if this should be taken into account. 

%%%%%%%%%%%%%%%%%%%%%%%%%%%%%%%%%%%%%%%%%%%%%%%%%%

% Don't change these lines
\bsp	% typesetting comment
\label{lastpage}
\end{document}